\documentclass[12pt]{article}

\usepackage{amsmath, amsthm, amsfonts, amssymb, bm, bbm}
\usepackage{algorithm, algorithmic, graphicx,subfigure,epstopdf}
\usepackage{hyperref, xcolor,tabularx,booktabs,enumerate,natbib,xr}
\usepackage{tikz}
\usetikzlibrary{arrows.meta, positioning, fit}
\usepackage{hyperref}
\hypersetup{colorlinks,linkcolor={blue},citecolor={blue},urlcolor={red}}

\DeclareMathOperator*{\argmax}{arg\,max}

\newcommand{\greekbold}[1]{\mbox{\boldmath $#1$}}
\newcommand{\betabf}{\greekbold{\beta}}
\newcommand{\bx}{\bm x}
\newcommand{\by}{\bm y}
\newcommand{\bz}{\bm z}
\newcommand{\bB}{\bm B}
\newcommand{\bbR}{\mathbb R}
\newcommand{\cO}{O}
\newcommand{\co}{o}

\def\eop{\hfill {\large $\Box$}}

\newtheorem{corollary}{{\bf Corollary}}
\newtheorem{assumption}{{\bf Assumption}}
\newtheorem{definition}{{\bf Definition}}
\newtheorem{theorem}{{\bf Theorem}}

\newtheorem{example}{{\bf Example}}

\begin{document}

\title{\Large{\textbf{Structural Classification of Locally Stationary Time Series Based on Second-order Characteristics}}}
\author{
\bigskip
Chen Qian$^{1}$ \thanks{Email: mrcqian@ucdavis.edu.}, Xiucai Ding$^{1}$\thanks{Email: xcading@ucdavis.edu.}, and Lexin Li$^{2}$ \thanks{Email: lexinli@berkeley.edu.} \\
\normalsize{\textit{$^1$University of California at Davis}} \\
\normalsize{\textit{$^2$University of California at Berkeley}}
}
\date{}
\maketitle

\baselineskip=20pt
\begin{abstract}
Time series classification is crucial for numerous scientific and engineering applications. In this article, we present a numerically efficient, practically competitive, and theoretically rigorous classification method for distinguishing between two classes of locally stationary time series based on their time-domain, second-order characteristics. Our approach builds on the autoregressive approximation for locally stationary time series, combined with an ensemble aggregation and a distance-based threshold for classification. It imposes no requirement on the training sample size, and is shown to achieve zero misclassification error rate asymptotically when the underlying time series differ only mildly in their second-order characteristics. The new method is demonstrated to outperform a variety of state-of-the-art solutions, including wavelet-based, tree-based, convolution-based methods, as well as modern deep learning methods, through intensive numerical simulations and a real EEG data analysis for epilepsy classification. 
\end{abstract}

\noindent{\bf Key Words:} 
Autocovariance function; Autoregressive approximation; Electroencephalogram; Locally stationary time series; Time series classification.

\baselineskip=20pt

\section{Introduction}

Classification of time series is a fundamental and widely studied problem, with applications in fields ranging from finance to biomedical analysis. Our motivation stems from epilepsy  classification, where the goal is to distinguish seizure-related patterns from non-seizure brain activity using electroencephalography (EEG), a non-invasive technology that records brain's electrical activity over time with high temporal resolution. As demonstrated in recent studies \citep{Jing2023Development, Kiessner2024Reaching}, automated methods, including those based on classical time series models and deep learning models, have shown promise in enabling reliable EEG pathological pattern classification, which in turn helps alleviate the substantial clinical burden associated with the time-consuming and expertise-intensive process of manual EEG interpretation. Despite recent progress, however, numerous challenges remain, including non-stationarity of the EEG signals, high inter-subject variability, the inherently low signal-to-noise ratio of EEG recordings, and the severe class imbalance due to the limited number of seizure patients relative to health controls. In this article, we focus on the structural classification problem of \emph{locally stationary} time series; namely, given two sets of time series, each associated with a distinct class, e.g., seizure versus non-seizure, we aim to learn interpretable and discriminative features from complex and noisy time series that can help accurately assign a new time series to the correct class. 

There is an extensive literature on the classification of \emph{stationary} time series, often referred to as time series discriminant analysis; see the monographs, \citet{shumway2000time} and \citet{maharaj2019time}, for comprehensive reviews. In contrast, much less is known about the classification of locally stationary time series, where the underlying structures evolve over time. Existing approaches for \emph{locally stationary} time series can be broadly classified into two main categories. The first category is rooted in traditional time series models, which typically focus on extracting frequency-domain features. Examples include the methods based on the locally stationary wavelet process \citep{nason2000wavelet, fryzlewicz2009consistent,krzemieniewska2014classification}, the one based on the smooth localized complex exponential model \citep{huang2004discrimination}, the one based on the spectral density under Gaussian distribution \citep{sakiyama2004discriminant}, the one based on stochastic cepstra and log-spectra \citep{krafty2016discriminant}, and the ones based on the variance of discrete wavelet transform \citep{maharaj2007discrimination}, among others. Under certain model assumptions, such as the locally stationary wavelet process, this family of solutions often come with theoretical guarantees. We also note that, for a special class of autoregressive models, \citep{chandler2006discrimination} proposed a discriminant procedure based on the shape of the residual variance functions. The second category relies on machine learning and deep learning methods, which are highly flexible and can accommodate various types of time series data. Examples include the methods that extract summary statistics as features then combine with tree-based methods for classification \citep{lines2012shapelet, baydogan2013bag, deng2013time, hills2014classification, baydogan2016time, cabello2020fast}, the ones that transform time series using random convolutional kernels followed by some linear classifiers \citep{dempster2020rocket, tan2022multirocket, jacques2022analysing, dempster2023hydra}, the ones that directly compare distances between time series through dynamic time warping and its variants \citep{marteau2008time, jeong2011weighted, gorecki2013using, lines2015time, Kate2016}, and the ones that employ deep neural networks for feature extraction then the final layer for classification \citep{ismail2019deep, sherstinsky2020fundamentals}. This category of solutions is designed for generic time series and may overlook specific underlying temporal structures, and as a result, the feature construction can be ad hoc and may lack theoretical justification. 

In this article, we propose a new approach for extracting interpretable and discriminative features based on \emph{time-domain}, \emph{second-order} characteristics, with the goal of classifying two groups of \emph{locally stationary} time series. Our proposal is built upon the structural approximation that any short-range dependent locally stationary time series can be well approximated by a time-varying autoregressive (AR) process, where the order is adaptive to the temporal dependence decay and the length of the time series \citep{ding2023autoregressive}. This approximation provides a principled basis for our classification task. We construct our discriminative features based on the corresponding time-varying AR coefficient functions from this approximation. We then propose an ensemble step to aggregate the extracted features, and develop a distance-based threshold for the aggregated feature to distinguish the two classes of time series. We rigorously establish the theoretical justification and guarantee for our method. We systematically compare with a large number of alternative solutions, and demonstrate the superior empirical performance of our method. 

More specifically, as we show in Theorem \ref{thm_1}, the second-order characteristics of the time series, including local stationarity, smooth structural variation, and temporal decay rates, can be fully captured by the corresponding time-varying AR coefficient functions. Then, as we show in Theorem \ref{thm_2}, if the two classes of time series differ in their second-order properties in terms of their time-varying autocovariance functions \eqref{eqn_covij}, these differences would be reflected in those AR coefficient functions as well. In other words, a class of time series exhibiting greater nonstationarity tends to produce more fluctuating AR coefficients. This motivates us to construct the discriminative features in the form of a maximum deviation of those AR coefficient functions. Consequently, our extracted features are interpretable, as they directly reflect second-order characteristics in the time domain. This interpretability is particularly evident in our EEG data analysis in Section \ref{sec:realdata}, where the clinically abnormal group displays more pronounced nonstationary fluctuations in these features compared to the normal group.

Given the observed time series from the two classes, we employ the methods of sieves and ordinary least squares (OLS) to estimate the AR coefficient functions as in \eqref{eq_phihat}. Such estimators are consistent under some mild smoothness assumptions on the autocovariance functions. We then construct a sequence of lag-based features for each time series as in \eqref{eq_maxdeviation}, by measuring the maximum deviation of the estimated AR coefficients, which are closely tied to the second-order characteristics of the time series through their autocovariance structures. Since the order of the AR approximation is theoretically adaptive to both the temporal dependence decay and the length of the time series, different series may require different AR orders to achieve optimal approximation. Accordingly, we further develop a cross-validation-based approach to numerically select both the order $b$ and the number of basis functions $c$ in the AR approximation of each individual time series. 

Next, we propose an ensemble step that aggregates the features, by taking a maximum over lags as in \eqref{maximu_feature_D}, for each individual time series. This is motivated by the observation that, if we include all constructed features for classification, it would increase the computational burden and also introduce additional variation due to the data-driven selection of the approximation order. At the high level, this ensemble approach prioritizes those features derived from AR coefficients with larger lags. This is also motivated by our theoretical result in Theorem \ref{thm_1}, which shows that the higher-order AR features can be interpreted as linear weighted averages of a slowly diverging number of autocovariance functions. As a result, even mild differences in the autocovariance structures between the two classes can be amplified through these summarized features, thereby yielding highly discriminative representations for classification, as we theoretically justify in Theorem \ref{thm_2}. 

Finally, we develop a distance-based method to select the threshold as in \eqref{equa_tau} that distinguishes between the two classes of time series. As we show in Corollary \ref{cor_1}, the resulting classification rule achieves consistent classification under mild conditions. 

We compare our method with a large number of alternative solutions, both analytically in Section \ref{subsec:comparison} and numerically in Section \ref{sec:simulations}. Moreover, we remark that, our method is built upon a key result of \citet{ding2023autoregressive} but otherwise differs significantly from \citet{ding2023autoregressive} in terms of the goal and focus. \citet{ding2023autoregressive} established a fundamental result for locally stationary time series, i.e., any locally stationary time series can be well approximated by a locally stationary AR process, but did not target time series classification. We build on this AR approximation, develop a classification solution, and establish its theoretical guarantees. 

In summary, our proposed method is structurally grounded, theoretically justified, computationally efficient, and practically competitive. It thus makes a useful and novel contribution to the toolbox of  time series classification. 

Throughout this article, we adopt the following notation. For two sequences of real numbers $\{a_n\}$ and $\{b_n\}$, $a_n = \cO(b_n)$ denotes that $|a_n| \leq C|b_n|$ for some finite constant $C > 0$, $a_n = \co(b_n)$ denotes that $a_n \le c_nb_n$ for some positive sequence $c_n$ that approaches zero. For a random variable $x$ and some constant $q \geq 1$, $\|x\|_q = \left(\mathbb{E}|x|^q\right)^{1/q}$ denotes its $L_q$ norm. For a sequence of random variables $\{x_n\}$ and positive real values $\{a_n\}$, $x_n = \cO_{L_q}(a_n)$ denotes that $x_n / a_n$ is bounded in $L_q$ norm, i.e., $\|x_n / a_n\|_q \leq C$ for some finite constant $C$. Similarly, we can define $x_n = \co_{L_q}(a_n)$. 

The rest of the article is organized as follows. Section \ref{sec:method} develops our feature construction and classification procedure, and compares with numerous alternative solutions. Section \ref{sec:theory} presents the theoretical justification and guarantee for our proposed method. Section \ref{sec:simulations} conducts intensive simulations. Section \ref{sec:realdata} revisits the motivating EEG classification problem. The supplementary appendix collects all technical proofs and additional numerical results.

\section{Feature Construction and Classification}
\label{sec:method}

\subsection{Feature construction based on AR approximation}
\label{subsec:feature}

We first consider a locally stationary time series $\bz = (z_i)_{i=1}^{n} \in \bbR^{n}$ from a single subject, where $n$ denotes the length of the time series. To simply our presentation, we assume throughout that the time series have mean zero. Later, we briefly remark on how to extend our method to handle nonzero mean. Following \citet{ding2023autoregressive}, there exists an autoregressive (AR) approximation for $\bz$, in that, for $i > b$, 
\begin{equation}\label{eq_firstapproximation}
z_i=\sum_{j=1}^b \phi_j\left( \frac{i}{n} \right) z_{i-j} + \epsilon_i + \co_{L_2}(1),
\end{equation}
where $\phi_j(t)$, $j = 1, \ldots, b$, $t \in [0,1]$, are a set of smooth AR coefficient functions, $b$ denotes the number of such functions, and $\{\epsilon_i\}$ is a time-varying white noise process. The value of $b$ depends on the time series length $n$, and specifies the \emph{order} of the AR approximation. It is usually much smaller than $n$ when $\bz$ has a short-range temporal dependence.

Meanwhile, $\phi_j(t)$ can be further approximated using a slowly divergent number of basis functions \citep{chen2007large}, 
\vspace{-0.01in}
\begin{equation}\label{eq_secondapproximation}
\phi_j(t) = \sum_{\ell=1}^{c} a_{j \ell}\alpha_{\ell}(t) + \cO(c^{-d}), \;\; j = 1, \ldots, b, \; t \in [0, 1],
\end{equation}
where $\alpha_\ell(t)$, $\ell = 1, \ldots, c$, are a set of pre-chosen sieve basis functions defined on $[0, 1]$, $c$ denotes the number of basis functions, $a_{j\ell}$'s are the coefficients, and $d$ represents the smoothness of $\phi_j(t)$. Some commonly used sieve basis functions include the normalized Fourier basis, orthonormal polynomials, Daubechies wavelets, and spline functions.

Together, under mild conditions \citep[Eq.(3.8)]{ding2023autoregressive}, for $i > b$, we have that, 
\begin{equation}\label{eq_third_ols}
z_i= \sum_{j=1}^b \sum_{\ell=1}^c a_{j \ell} \alpha_\ell \left( \frac{i}{n} \right) z_{i-j} + \epsilon_i + \cO_{L_2}\left(\frac{b^2}{n} + bc^{-d}\right).
\end{equation}
We employ OLS to obtain an estimate of the AR coefficient function $\phi_j(t)$ as,
\begin{equation}\label{eq_phihat}
\widehat \phi_j(t) = \widehat\betabf^\top \bB_j(t),
\end{equation}
where $\bB_j(t) \in \bbR^{bc}$ has $b$ blocks of elements, with the $j$th block being $(\alpha_1(t),...,\alpha_c(t))^\top \in \mathbb{R}^c$ and zeros otherwise, and $\widehat\betabf \in \mathbb{R}^{bc}$ is the OLS estimate of the coefficients $a_{j \ell}$'s organized as a $bc \times 1$ vector, where the ``response" is $(z_{b+1}, \ldots, z_n)^\top \in \bbR^{n-b}$, and the ``design matrix" has $(n-b)$ rows and $bc$ columns with the $\mathrm{s}$th row being $\big(z_{\mathrm{s}+b-1},\ldots,z_{\mathrm{s}} \big) \otimes \big(\alpha_1((\mathrm{s}+b)/n),\ldots, \alpha_c((\mathrm{s}+b)/n)\big), 1 \leq \mathrm{s} \leq n-b,$ where $\otimes$ is the Kronecker product.

Motivated by the observations that the AR coefficient functions $\phi_j(t), 1 \leq j \leq b,$ capture the second-order characteristics of the time series, and a time series exhibiting greater nonstationarity tends to produce more fluctuating $\phi_j(t)$, we take the maximum deviation of $\phi_j(t)$ as the discriminative feature for our classification task, i.e., 
\begin{equation}\label{eq_maxdeviation}
D(j) = \sup_{t_1, t_2 \in [0,1]} \left| \widehat{\phi}_j(t_1)-\widehat{\phi}_j(t_2) \right|.
\end{equation}

In the AR approximation, it is important to choose the order $b$ and the number of sieve basis functions $c$. In principle, for $b$, as shown in \citet{ding2023autoregressive}, it is adaptive to the temporal dependence decay rate. Moreover, $b$ is usually much smaller than $n$, if $\text{Cov}(z_i, z_j)$ decays sufficiently fast when $|i - j|$ becomes large. For $c$, it should be reasonably large so that the error term in \eqref{eq_secondapproximation} can be negligible. Meanwhile, it should not be too large so that the coefficients $a_{j\ell}$'s in \eqref{eq_third_ols} can be adequately estimated. In practice, we adopt the leave-one-out cross-validation approach of \cite{bishop2006pattern} to select $b$ and $c$. 

Next, we consider a set of locally stationary time series $\{ \bz_k \}_{1 \le k \le N}$ from $N$ subjects, each with the length $n_k$. Note that we do \emph{not} require each time series to have the equal length. For each time series $\bz_k \in \bbR^{n_k}$, we choose $b_k$ and $c_k$ using cross-validation. Again, the choice of $b_k$ does not have to be the same. We then propose an ensemble step to aggregate the discriminative features across all $N$ time series. Specifically, we compute 
\begin{equation}\label{minimal_b}
b^{*} = \mathrm{min}_{1 \le k \le N} b_k.
\end{equation}
Let $D_k(j)$ denote the maximum deviation feature in \eqref{eq_maxdeviation} for the $k$th time series $\bz_k$ at lag $j$. We construct the aggregated discriminative feature as  
\begin{equation}\label{maximu_feature_D}
S_k = \mathrm{sup}_{\max\{b_k - b^{*} +1, b^{*}\} \le j \le b_k} D_k(j), \;\;\; k = 1, \ldots, N.
\end{equation}
At the high level, this ensemble step prioritizes those features derived from AR coefficients with larger lags, which is motivated by our theoretical result in Theorems \ref{thm_1} and \ref{thm_2}, which essentially shows that even mild differences in the autocovariance structures between two classes of time series can be amplified through these features. As we show later in both theory and numerical analysis, such an essemble treatment allows us to obtain accurate classification.

\subsection{Threshold-based classification}
\label{subsec:classify}

We now consider two classes of time series, $\{\bx_{k_1}\}_{1 \leq k_1 \leq N_1}$ from class 1 of $N_1$ subjects with the $k_1$th series $\bx_{k_1}=(x_{k_1, i})_{i=1}^{n_{k_1}} \in \mathbb{R}^{n_{k_1}}$ of length $n_{k_1}$, $k_1=1, \ldots, N_1$, and $\{\by_{k_2}\}_{1 \leq k_2 \leq N_2}$ from class 2 of $N_2$ subjects with the $k_2$th series $\by_{k_2}=(y_{k_2, i'})_{i'=1}^{m_{k_2}} \in \mathbb{R}^{m_{k_2}}$ of length $m_{k_2}$, $k_2=1, \ldots, N_2$. We assume each time series is long, i.e., $n_{k_1}$ and $m_{k_2}$ are large, but each time series does not have to have the equal length. Moreover, we do not require the number of subjects $N_1, N_2$ to be large or balanced. This setting aligns well with biomedical applications such as EEG-based epilepsy classification, where the EEG series are long, but the number of patients can be limited. We assume $\bx_{k_1}$ and $\by_{k_2}$ are generated from locally stationary processes,
\begin{equation}\label{eq_G1G2} 
x_{k_1, i} \sim G_1(i/n_{k_1}, \mathcal{F}_{1i}), \quad\quad y_{k_2, i'} \sim G_2(i'/m_{k_2}, \mathcal{F}_{2i'}), 
\end{equation} 
where $\{\mathcal{F}_{1i}\}$ and $\{\mathcal{F}_{2i'}\}$ are independent filtrations representing the randomness within each time series, and $G_1(\cdot, \cdot)$ and $G_2(\cdot, \cdot)$ are two functions \citep{dahlhaus2019towards,Wu2005}. Our goal is to learn highly discriminative and interpretable class features related to the generating mechanisms $G_1$ and $G_2$ in \eqref{eq_G1G2} from the training data $\{\bx_{k_1}\}_{1 \leq k_1 \leq N_1}$ and $\{\by_{k_2}\}_{1 \leq k_2 \leq N_2}$, such that for a new time series $\bz$ from the testing data, we can accurately assign it to the correct class by comparing its extracted feature with the learned class features.

Following the construction procedure in Section \ref{subsec:feature}, we obtain the aggregated features
\begin{equation} \label{maximu_feature_D_xy}
S^{x}_{k_1} = \sup_{\max\{b^x_{k_1} - b_x^{*} + 1, b_x^{*}\} \le j \le b^x_{k_1}} D^{x}_{k_1}(j), \quad\quad
S^{y}_{k_2} = \sup_{\max\{b^y_{k_2} - b_y^{*} + 1, b_y^{*}\} \le j \le b^y_{k_2}} D^{y}_{k_2}(j), 
\end{equation}
where $b^{*}_x = \min_{1 \le k_1 \le N_1} b^x_{k_1}$, and $b^{*}_y = \min_{1 \le k_2 \le N_2} b^y_{k_2}$. We then obtain the summary features averaged across the subjects in each class as
\begin{equation}\label{medi_x_y}
\bar{S}_x = \mathrm{median}_{1 \le k_1 \le N_1} S_{k_1}^x, \quad\quad 
\bar{S}_y = \mathrm{median}_{1 \le k_2 \le N_2}S_{k_2}^y.
\end{equation}

We next derive a threshold value $\vartheta$ for classification. Specifically, we first obtain a grid of $M+1$ points on the interval $[C_1, C_2]$, where \begin{equation}\label{com_C}
C_1 = \frac{1}{2} \mathrm{min}_{1 \le k_1 \le N_1, 1 \le k_2 \le N_2} \left\{ S^{x}_{k_1}, S^{y}_{k_2} \right\}, \quad\quad
C_2 = 2 \mathrm{max}_{1 \le k_1 \le N_1, 1 \le k_2 \le N_2} \left\{ S^{x}_{k_1}, S^{y}_{k_2} \right\}.
\end{equation}
We then generate the set of candidate threshold values $\{\vartheta_i\}_{1 \le i \le M+1}$, where $\vartheta_i = C_1 + (C_2 - C_1)(i-1)/M$. Then, for each $\vartheta_i$, $i = 1, \ldots, M+1$, we compute the number of time series correctly assigned to the two classes, respectively, as
\begin{equation}\label{corr_N_1}
N^x_{i} = 
\begin{cases} 
\sum_{k_1 = 1}^{N_1} \mathbb{I}(S^{x}_{k_1} \le \vartheta_i), & \text{if } \bar{S}_x < \bar{S}_y, \\
\sum_{k_1 = 1}^{N_1} \mathbb{I}(S^{x}_{k_1} > \vartheta_i), & \text{if } \bar{S}_x > \bar{S}_y,
\end{cases}
\quad\quad
N^y_{i} = 
\begin{cases} 
\sum_{k_2 = 1}^{N_2} \mathbb{I}(S^{y}_{k_2} \le \vartheta_i), & \text{if } \bar{S}_x < \bar{S}_y, \\
\sum_{k_2 = 1}^{N_2} \mathbb{I}(S^{y}_{k_2} > \vartheta_i), & \text{if } \bar{S}_x > \bar{S}_y,
\end{cases}
\end{equation}
where $\mathbb{I}(\cdot)$ is the indicator function. We then choose the threshold value from $\{\vartheta_i\}_{1 \leq i \leq M+1}$ that maximizes the overall classification accuracy, in that
\begin{equation}\label{equa_tau}
\vartheta = \min \left\{  \argmax_{\vartheta_i, 1\le i \le M+1} \frac{ N^x_{i}+ N^y_{i}}{N_1+N_2} \right\}.
\end{equation}
We briefly remark that the choice of $C_1$ and $C_2$ ensures that the classification boundary lies within the candidate set $\{\vartheta_i\}_{1 \le i \le M+1}$. Moreover, $M$ should be sufficiently large to ensure that there are enough candidates near the boundary. If there are more than one candidates that  maximize the classification accuracy in \eqref{equa_tau}, we then select the smallest value among them. Our numerical experiments show that this approach works well empirically. 

\begin{figure}[t!]
\centering
\hspace{-50pt}
\includegraphics[width=18cm, height=8cm]{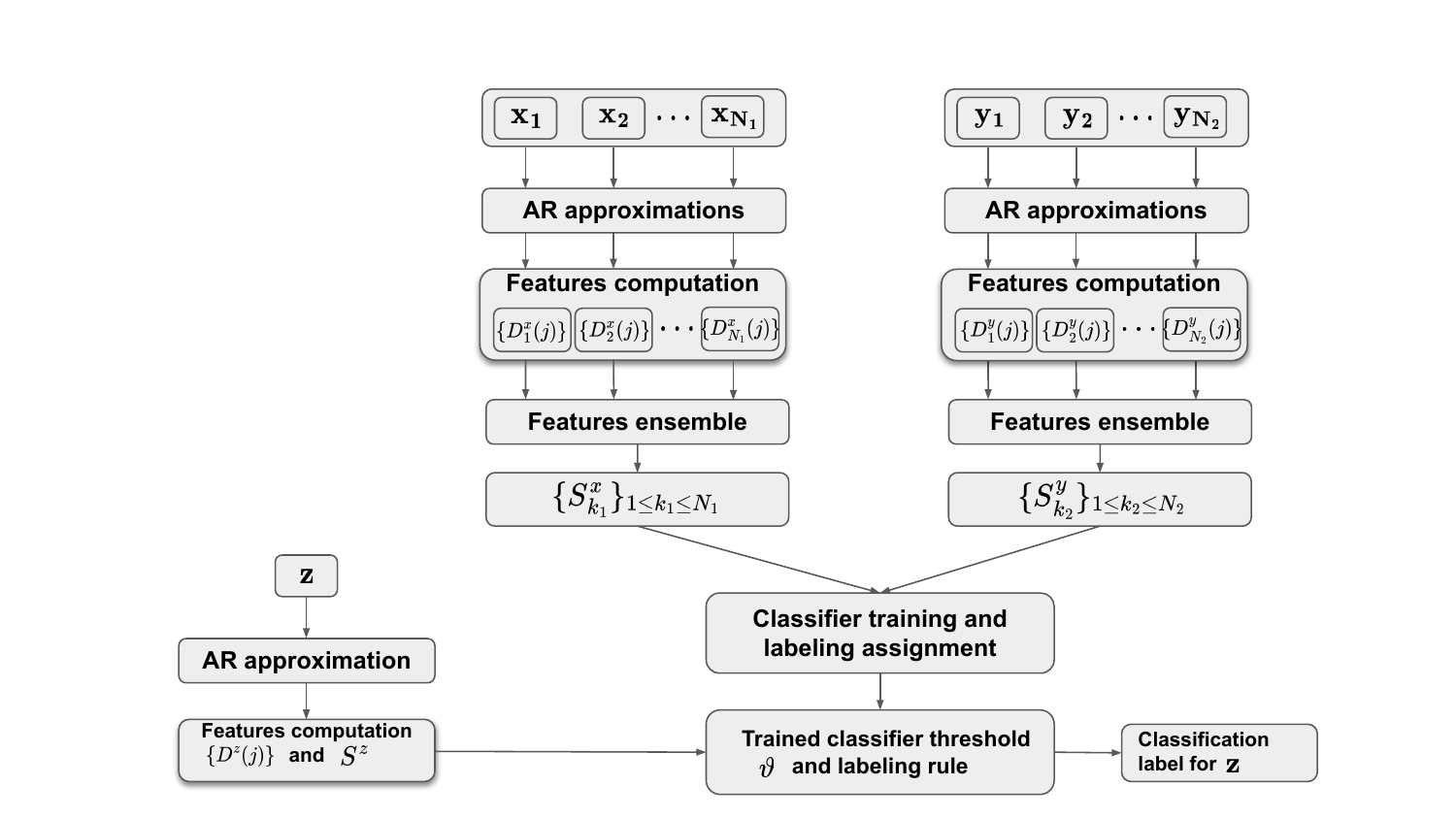}
\caption{Schematic diagram for the proposed binary structural classification procedure.}
\label{fig_plotone}
\end{figure}

Finally, given a new time series $\bz$ from the testing data, we compute the aggregated discriminative feature $S^z$ following \eqref{maximu_feature_D}. We then compare $S^z$ with the threshold value $\vartheta$. If $S^z < \vartheta$, we assign $\bz$ to class 1 when $\bar{S}_x < \bar{S}_y$ and to class 2 when $\bar{S}_x > \bar{S}_y$; Otherwise, if $S^z > \vartheta$, we assign $\bz$ to class 2 when $\bar{S}_x < \bar{S}_y$ and to class 1 when $\bar{S}_x > \bar{S}_y$. Algorithm \ref{algo_1} summarizes the above classification procedure, while Figure \ref{fig_plotone} presents a schematic plot. 

\begin{algorithm}[t!]
\caption{Binary structural classification procedure}
\label{algo_1}    
\textbf{Input:} The training time series $\{\bx_{k_1}\}_{1 \leq k_1 \leq N_1}$ and $\{\by_{k_2}\}_{1 \leq k_2 \leq N_2}$ with known labels, and the testing time series $\bz$ with unknown label, and the sieve basis functions.
\smallskip

\textbf{Step 1:} For the training time series $\{\bx_{k_1}\}_{1 \leq k_1 \leq N_1}$ and $\{\by_{k_2}\}_{1 \leq k_2 \leq N_2}$, compute the aggregated discriminative features $S^{x}_{k_1}, S^{y}_{k_2}$ based on \eqref{maximu_feature_D_xy}, and the average features $\bar{S}_x, \bar{S}_y$ based on \eqref{medi_x_y}.
\smallskip

\textbf{Step 2:} Compute the threshold value $\vartheta$ based on \eqref{equa_tau}.
\smallskip

\textbf{Step 3:} For the testing time series $\bz$, compute the aggregated discriminative feature $S^{z}$.
\smallskip

\textbf{Step 4:} Compare $S^{z}$ with the $\vartheta$ and carry out the classification: If $S^z < \vartheta$, assign $\bz$ to class 1 when $\bar{S}_x < \bar{S}_y$ and to class 2 when $\bar{S}_x > \bar{S}_y$; Otherwise, if $S^z > \vartheta$, assign $\bz$ to class 2 when $\bar{S}_x < \bar{S}_y$ and to class 1 when $\bar{S}_x > \bar{S}_y$.
\end{algorithm}

We make a few remarks regarding our proposed structural classification method. 

First, in the extreme case that $\bar{S}_x = \bar{S}_y$, our method would not be able to assign labels. This occurs when the two classes of time series exhibit nearly identical dynamics, and as such the AR approximation functions could not capture the difference. Nevertheless, as we will see in our theoretical justification as well as numerical experiments, as long as the dynamics of the two classes of time series differ slightly, $\bar{S}_x$ is to be different from $\bar{S}_y$ with a high probability, enabling our method to distinguish between the two classes. 

Second, our method has been designed for the case when at least one class of time series is nonstationary in correlation. If both classes are stationary in correlation, i.e., their means are constant and the autocorrelation functions (ACF) depend only on the lag, the corresponding AR coefficient functions become constants, and the maximum deviation features become zero. As such, our method is no longer applicable. In Appendix \ref{append:stationary}, we first develop a prescreening step to screen out such a special case. We then propose a modified classification method when both classes of time series are indeed stationary in correlation. 

Finally, our method so far has assumed that the time series has mean zero. When the mean is nonzero, we can add a smooth intercept $\phi_0(t)$ to \eqref{eq_firstapproximation}, such that 
\begin{equation*}
z_i = \phi_0\left( \frac{i}{n} \right) + \sum_{j=1}^b \phi_j\left( \frac{i}{n} \right) z_{i-j} + \epsilon_i + \co_{L_2}(1).
\end{equation*}
The intercept term $\phi_0 (t)$ can be approximated similarly using \eqref{eq_secondapproximation}. Accordingly, we can update \eqref{eq_third_ols} and \eqref{eq_phihat}, and construct the maximum deviation measure as in \eqref{eq_maxdeviation}, for all $j = 0, \ldots, b$. For two classes of time series with nonzero means, they may differ in their first-order characteristics, i.e., the mean functions, only. As such, we propose to first do classification based on the means, before applying our Algorithm \ref{algo_1}. Specifically, we first compute the mean function of each time series in the two classes of training samples, then average these mean functions within each class to obtain the class-wise average. Next, we compute the mean function of the unlabeled time series, and compute its $L_{\infty}$ norm distance to each class-wise average. If this difference is sufficiently large, we assign the label based on the proximity in this norm. Otherwise, we proceed with Algorithm \ref{algo_1} for classification beyond the mean.

\subsection{Comparison with alternative methods}
\label{subsec:comparison}

We compare our proposed time series classification method with a number of existing solutions, first analytically in this section, then numerically in Section \ref{sec:simulations}. We focus on the following ten alternative methods, grouped into five categories. 

The first category is functional logistic regression \citep[FLogistic,][]{jacques2022analysing}, which treats each time series as a functional curve, then fits a logistic model.

The second category is wavelet-based, including spectral discriminant analysis based on locally stationary wavelet model \citep[LSW,][]{fryzlewicz2009consistent,  krzemieniewska2014classification}, and discrete wavelet transform \citep[DWT,][]{maharaj2007discrimination}. The former models time series using locally stationary wavelet process and extracts the evolutionary wavelet spectrum as features for classification. It then computes the squared quadratic distance between the empirical wavelet spectrum of the unlabeled time series and the class-wise spectrum, and assigns the label based on the smaller distance. The latter divides each time series into consecutive blocks, computes the discrete wavelet transform coefficients for each block, and extracts the maximum overlap discrete wavelet transform variance as features for classification. It then applies quadratic discriminant analysis. 

The third category is tree-based, including supervised time series random forest \citep[STSForest,][]{cabello2020fast}, and shapelet forest \citep[Shapelet,][]{lines2012shapelet}. The former incorporates the original time series, its first-order differences, and frequency-domain representations as features for classification, while the latter measures the similarity between time series using shape-based representations. For classification, both methods train random forests using the extracted features. 

The fourth category is deep neural networks-based for time series data, most notably, convolutional neural networks \citep[TimeCNN,][]{ismail2019deep}, and recursive neural networks \citep[TimeRNN,][]{sherstinsky2020fundamentals}. The former employs convolutional filters to extract hierarchical features from the time series, while the latter captures sequential dependency through recurrent connections. Both methods perform classification by directly applying the sigmoid function to their output values. 
 
The fifth category is convolution-based, including Rocket \citep{dempster2020rocket}, MultiRocket \citep{tan2022multirocket}, and Arsenal \citep{dempster2023hydra}. This family of methods begin by convoluting each time series with a large number of randomly generated kernels, producing a set of transformed time series. Rocket then extracts two summary statistics per kernel from each transformed series as features for classification, while MultiRocket computes three additional features per kernel. Both methods then fit an $L_2$ regularized linear classifier using the extracted features. Arsenal is an ensemble method that aggregates the outputs of multiple Rocket classifiers. 

In contrast, our proposed method constructs the discriminative features by explicitly capturing the second-order characteristics of time series in the time domain. The resulting features are directly interpretable, and are highly sensitive to even slight difference of the second-order dynamics of the two classes of time series.

\section{Theoretical Analysis}
\label{sec:theory}

\subsection{Locally stationary time series and population feature analysis}
\label{subsec:ar-approx}

We first show in Theorem \ref{thm_1} that the second-order characteristics of a locally stationary time series can be well captured by the maximum deviation feature in \eqref{eq_maxdeviation}. 

We begin by providing some background on locally stationary time series. We focus on a general class of locally stationary time series following the definition introduced in \cite{ding2023autoregressive}, which covers many commonly used locally stationary time series models in the literature. See for instance \cite{DZ1,dahlhaus2019towards,dahlhaus2012locally,RSP,KPF,DPV,dette2020prediction,MR3097614,WZ1}.

\begin{definition}\label{def_1}
A non-stationary time series $\bz = (z_i)_{i=1}^{n}$ is a locally stationary time series, if there exists a function $\gamma(t, k) : [0,1] \times \mathbb{N} \to \mathbb{R}$, such that 
\begin{equation}\label{eqn_covij}
\operatorname{Cov}(z_i, z_j) = \gamma(t_i, |i - j|) + \cO\left(\frac{|i - j| + 1}{n}\right), \quad t_i = \frac{i}{n}.
\end{equation}
Moreover, suppose $\gamma$ is Lipschitz continuous in $t$, and for any fixed $t \in [0, 1]$, $\gamma(t, \cdot)$ is the autocovariance function of a stationary process.
\end{definition}

We adopt the physical representation for locally stationary time series \citep{Wu2005}, and assume that $\bz = (z_i)_{i=1}^{n}$ is generated according to 
\begin{equation}\label{model_1}
z_i = G\left( i/n, \mathcal{F}_i \right), 
\end{equation}
where $\mathcal{F}_i = (\ldots, \xi_{i-1}, \xi_i)$ whose elements are i.i.d.\ random variables, and $G:[0,1] \times \mathbb{R}^\infty \to \mathbb{R}$ is a measurable function, such that $\zeta_i(t) = G\left(t, \mathcal{F}_i\right)$ is a properly defined random variable for all $t \in [0,1]$. Moreover, it satisfies that, for some $q > 2$ and $C > 0$,
\begin{equation}\label{eqn_assumpG}
\| G(t_1, \mathcal{F}_i) - G(t_2, \mathcal{F}_i) \|_q \leq C \|t_1 - t_2\|, \quad\quad 
\sup_{t \in [0,1]} \max_{1 \leq i \leq n} \| G(t, \mathcal{F}_i) \|_q < \infty.
\end{equation}
We note that under \eqref{model_1}, the autocovariance function in \eqref{eqn_covij} can be explicitly written as 
\begin{equation}\label{eq_autocovariancefunctionfunction}
\gamma(t,j)=\operatorname{Cov}(G(t, \mathcal{F}_0), G(t, \mathcal{F}_j)). 
\end{equation}

To quantify the short-range dependence of the locally stationary time series in (\ref{model_1}), we employ the physical dependence measure \citep{dahlhaus2019towards,Wu2005}, which offers a convenient tool to study the temporal dependence of time series. 

\begin{definition}[Physical dependence measure]\label{defn_physcialdependence} 
Suppose (\ref{eqn_assumpG}) holds for some $q>2$. For $j \geq 0$, the physical dependence measure of $\bz = (z_i)_{i=1}^{n}$ in (\ref{model_1}) is defined as
\begin{equation*}
\delta(j, q)=\sup_t \| G(t, \mathcal{F}_0)-G(t, \mathcal{F}_{0,j}) \|_q, 
\end{equation*}
where $\mathcal{F}_{0,j}=(\mathcal{F}_{-j-1}, \xi_{-j}', \xi_{-j+1}, \cdots, \xi_0)$, and $\{\xi_i'\}$ is an i.i.d.\ copy of $\{\xi_i\}$. For $j<0$, denote $\delta(j,q)=0$.
\end{definition}   

Next, we introduce the following assumption. 

\begin{assumption}\label{assu_basicassumption} 
For the mean zero locally stationary time series $\bz = (z_i)_{i=1}^{n}$ in (\ref{model_1}), suppose
\begin{equation}\label{eq_temporaldecay}
\delta(j,q) \leq C j^{-\tau}, \ \text{for all} \ j \geq 1,
\end{equation}
for some constants $\tau>1$ and $C>0$. Moreover, suppose 
\begin{equation}\label{eq_eigenvalueslowerbound}
\inf_{t \in [0,1]} \lambda_{\min}(\Gamma^j(t)) \geq \kappa,
\end{equation}
for some constant $\kappa>0$ and for all $j \geq 1$, where $\Gamma^j(t) = (\Gamma_{\ell_1 \ell_2}^j(t)) \in \mathbb{R}^{j \times j}$ is a $j \times j$ symmetric matrix whose entries are 
\begin{equation}\label{eq_gammaentries}
\Gamma_{\ell_1 \ell_2}^j(t)=\gamma(t, |\ell_1-\ell_2|),  \ \ell_1, \ell_2 = 1, \ldots, j,
\end{equation}
and $\lambda_{\min}(\Gamma^j(t))$ denotes the smallest eigenvalue of the matrix $\Gamma^j(t)$. 
\end{assumption}

We make two remarks regarding this assumption. First, the condition in (\ref{eq_temporaldecay}) indicates that the temporal dependence of ${z_i}$ decays polynomially, i.e., $\sup_i|\operatorname{Cov}(z_i, z_{i+j})| = \cO(j^{-\tau})$. A larger value of $\tau$ corresponds to a faster rate of decay. This condition encompasses exponential decay as a special case when $\tau = \infty$. We note that our algorithm does not require any specific value of $\tau$. However, for theoretical guarantees, we require $\tau$ to be relatively large, a condition that can be verified in practice \citep[][Remark A.2]{ding2024partial}. Second, the condition in (\ref{eq_eigenvalueslowerbound}) requires that the smallest eigenvalue of the time series covariance matrix to be bounded away from zero, which is a standard condition in the statistical literature on covariance or precision matrix estimation to prevent erratic behavior in the time series \citep{Yuan2010, chen2013,DZ1, ding2023autoregressive}. Equivalently, for a sufficiently large $n$, $\lambda_{\min}(\operatorname{Cov}(z_{1}, \ldots, z_{n})) \geq \kappa$. Moreover, \citet[][Proposition 2.9]{ding2023autoregressive} shows that, for short-memory locally stationary time series, the condition is equivalent to the time-varying spectral density function being uniformly bounded away from zero across all times and frequencies. In practice, this condition can be assessed through spectral analysis. The spectral density can be estimated using wavelet-based methods, such as those proposed in \cite{fryzlewicz2006haar, GN, nason2000wavelet}. Under (\ref{eq_eigenvalueslowerbound}), $\Gamma^b(t)$ is invertible for all $t \in [0,1].$ We denote its precision matrix as 
\begin{equation}\label{eq_precision}
\Omega^b(t) = \left( \Gamma^b(t) \right)^{-1} = \left( \Omega^b_{\ell_1 \ell_2}(t) \right) \in \mathbb{R}^{b \times b}.
\end{equation}

We now proceed to state our main results. Denote the population version of (\ref{eq_maxdeviation}) as 
\begin{equation}\label{eq_populationfeatures}
D^*(j) = \sup_{t_1, t_2 \in [0,1]}| \phi_j\left( t_1\right)-\phi_j\left( t_2\right)|, \;\; \text{ for } \; j = 1, \ldots, b. 
\end{equation}

\begin{theorem}\label{thm_1}
Suppose Assumption \ref{assu_basicassumption} holds. Given some large value of $b \equiv b(n)>0$, we have the following results hold.
\begin{enumerate}[(a)]
\item For $j = 1, \ldots, b$, 
\begin{equation*}
D^*(j)=\sup_{t_1, t_2 \in [0,1]}\left| \sum_{j'=1}^b \Omega^b_{j'j}(t_1) \gamma(t_1,j')- \sum_{j'=1}^b \Omega^b_{j'j}(t_2) \gamma(t_2,j') \right|.
\end{equation*}

\item For $j = 1, \ldots, b$, there exist a sequence $\{w_j(j')\}_{1 \leq j' \leq b}$ satisfying that 
\begin{equation}\label{eq_wjkdefinition}
|w_j(j')|=\cO \left((b/\log b)^{-\tau+1}+ \mathbb{I}(|j-j'|<b/\log b) r^{|j-j'| \log b/b} \right), 
\end{equation}
for all $j' = 1, \ldots, b$, where $0<r<1$ is some constant, so that
\begin{equation*}
D^*(j)=\cO\left(\sum_{j'=1}^b |w_j(j')|\sup_{t_1, t_2 \in [0,1]} \left| \gamma(t_1,j')-\gamma(t_2,j') \right|+ \sum_{j'=1}^b (j')^{-\tau} \sup_{t_1, t_2 \in [0,1]} \left| \Omega^b_{j'j}(t_1)- \Omega^b_{j'j}(t_2) \right| \right). 
\end{equation*}
Moreover, for $j \asymp b^{a}$ with some $0 < a \leq 1$,  
\begin{equation}\label{eq_bounbounbound}
D^*(j)=\cO\left(\sum_{j'=1}^b w_j(j')\sup_{t_1, t_2 \in [0,1]} \left| \gamma(t_1,j')-\gamma(t_2,j') \right|+b^{(-\tau+1)\alpha}+b^{-\tau+2} \right). 
\end{equation}

\item For $j \asymp b^{a}, 0 < a \leq 1$, there exist a sequence  $\{w_j(j')\}_{1 \leq j' \leq b}$ satisfying (\ref{eq_wjkdefinition}), some constants $c_j > 0$, and $E_j = \cO(b^{(-\tau+1)\alpha}+b^{-\tau+2}) > 0$, so that  
\begin{equation}\label{eq_bounbounbound11}
D^*(j) \geq c_j \sup_{t_1, t_2 \in [0,1]}\left|\sum_{j'=1}^b w_j(j')(\gamma(t_1,j')-\gamma(t_2,j')) \right| - E_j.
\end{equation}
\end{enumerate}    
\end{theorem}

We give the proof in Appendix \ref{append:proofthm1}. We make several remarks. Theorem \ref{thm_1}(a) states that the population discriminative feature in (\ref{eq_populationfeatures}) can be fully characterized through certain functional forms of the maximum deviation across its autocovariance function (\ref{eq_autocovariancefunctionfunction}), which summarizes the fluctuation of the second-order characteristics. Theorem \ref{thm_1}(b) provides an upper bound for $D^*(j)$, in terms of the maximum deviation of $\gamma(t,\cdot)$, and the precision matrix, which itself is a highly nonlinear function of $\gamma(t,\cdot)$. On the one hand, this bound is sharp for a stationary time series, where $\gamma(t, j') \equiv \gamma(j')$ and $\Omega_{j'j}^b(t) \equiv \Omega_{j'j}^b$, making the right-hand-side of the bound vanish. On the other hand, for each lag $j = 1 \ldots, b$, the first term of this bound is dominated by those lags close to $j$. When $j$ is large in the sense that $j \asymp b^{a}$, the second term of this bound is dominated by the first, as shown in (\ref{eq_bounbounbound}). As a result, $D^*(j)$ can be fully governed by a linear form of the maximum deviations across the autocovariance functions. Theorem \ref{thm_1}(c) provides a lower bound in the form of a weighted average of the autocovariance functions.

\subsection{Asymptotic guarantees of the proposed method}
\label{subsec:guarantee}

We next establish the asymptotic guarantee of our proposed classification method, in that the misclassification rate approaches zero with a high probability. 

Our procedure relies on the feature constructed in (\ref{eq_maxdeviation}), which is the sample analog of $D^*(j)$ based on the estimator in (\ref{eq_phihat}). Since this estimator is built on the series expansion in (\ref{eq_secondapproximation}), certain smoothness condition is required. As shown later in the proof, the necessary smoothness condition can be inherited from the properties of the autocovariance function that characterizes the second-order properties of the time series, as stated below. 

\begin{assumption}\label{assu_mainmainmain} 
Suppose the training time series $\{\bx_{k_1}\}_{1 \leq k_1 \leq N_1}$ and $\{\by_{k_2}\}_{1 \leq k_2 \leq N_2}$ are generated following (\ref{eq_G1G2}), and $G_1, G_2$ satisfy Definition \ref{def_1} and Assumption \ref{assu_basicassumption} with the corresponding parameters $\tau_1$ and $\tau_2$ as specified in (\ref{eq_temporaldecay}). Moreover, denote the autocovariance functions associated with $G_1$ and $G_2$ as $\gamma_1$ and $\gamma_2$, respectively.  Suppose there exist some integers $d_g \ge 1$, such that $\gamma_g(t, \cdot) \in C^{d_g}([0, 1])$ with respect to $t$, where $C^{d_g}([0, 1])$ denotes the function space on $[0, 1]$ of continuous functions that have continuous first $d_g$ derivative, $g=1,2$. 
\end{assumption}

We now proceed to state our main results. For the basis functions used in (\ref{eq_secondapproximation}), we denote
\begin{equation}\label{eq_basicquantity}
\xi_c = \sup_{t \in [0,1]} \sup_{1 \leq \ell \leq c} |\alpha_\ell(t)|, \quad\quad \zeta_c = \sup_{t \in [0,1]} \sqrt{\sum_{\ell=1}^c \alpha_\ell^2(t)}.  
\end{equation}
Since our procedure and theoretical results do not rely on the sample size of the training subjects, to simplify our presentation, we let $N_1 = N_2 = 1$. That is, the time series $\bm{x}$ is from class one, and $\bm{y}$ is from class two. To further simplify the notation, for the two time series, denote their lengths as $n_1$ and $n_2$, and the corresponding maximum deviation feature as $D_1(j)$ and $D_2(j)$ under the parameters $b_1,c_1$ and $b_2,c_2$. In addition, denote the population version as $D^*_1(j)$ and $D^*_2(j)$, respectively.  Moreover, for some constant $0 < a_g \leq 1,$ we choose some $b_g^*< b_g$ such that $b_g^* \asymp b_g^{a_g}, g=1,2$. Similar to (\ref{maximu_feature_D_xy}), we denote the feature of the two classes of time series as 
\begin{equation}\label{eq_featurefinaldefinition}
S_g:=\sup_{\max\{b_g-b_g^*+1, b_g^*\} \leq j \leq b_g} D_{g}(j). 
\end{equation} 

\begin{theorem}\label{thm_2}
Suppose Assumption \ref{assu_mainmainmain} holds with $N_1=N_2=1$. Moreover, suppose
\begin{equation}\label{eq_errorprecisionbound}
b_1c_1 \left(\frac{\xi_{c_1}^2}{\sqrt{n_1}}+\frac{\xi_{c_1}^2 {n_1}^{\frac{2}{\tau_1+1}}}{{n_1}} \right)+b_2c_2 \left(\frac{\xi_{c_2}^2}{\sqrt{n_2}}+\frac{\xi_{c_2}^2 n_2^{\frac{2}{\tau_2+1}}}{n_2} \right)=\co(1). 
\end{equation}
Then, for sufficiently large $n_1$ and $n_2$, we have the following results hold.
\begin{enumerate}[(a)]
\item For the two classes of time series $g=1,2$,
\begin{equation*}
\sup_{1 \leq j \leq b_g} |D_g(j) - D_g^*(j)|=\cO_{\mathbb{P}}\left(\xi_{c_g} \zeta_{c_g} \sqrt{\frac{b_g c_g}{n_g}} \left[1+b_g c_g^{-d_g}+\frac{b_g^2}{n_g} \right]+c_g^{-d_g}\right). 
\end{equation*}

\item For $g=1,2,$ denote 
\begin{equation*}
\Psi_g = b_g^{(-\tau_g+1) a_g}+b_g^{-\tau_g+2}+\xi_{c_g} \zeta_{c_g} \sqrt{\frac{b_g c_g}{n_g}} \left[1+b_g c_g^{-d_g}+\frac{b_g^2}{n_g} \right]+c_g^{-d_g}.
\end{equation*}
Suppose, for $\max\{b_g-b_g^*+1, b_g^*\} \leq j_g \leq b_g$, 
\vspace{-0.01in}
\begin{equation}\label{eq_boundoneboundtwo}
\sup_{t_1, t_2 \in [0,1]}\left|\sum_{j'=1}^{b_g} w_{j_g,g}(j')\left\{ \gamma_g(t_1,j')-\gamma_g(t_2,j') \right\} \right| \gg \Psi_g, \ g=1,2. 
\end{equation}
Furthermore, suppose, for $\max\{b_g-b_g^*+1, b_g^*\} \leq j_g \leq b_g$, and for $g_1 \neq g_2 \in \{1,2\}$ and some sufficiently large constant $C>0$,
\begin{align}\label{eq_assumtiondasdsadasdads}
\begin{split}
\sup_{t_1, t_2 \in [0,1]} & \left|\sum_{j'=1}^{b_{g_1}} w_{j_{g_1},g_1}(j')(\gamma_{g_1}(t_1,j')-\gamma_{g_2}(t_2,j')) \right| \\
& > C \sum_{j'=1}^{b_{g_2}} |w_{{j_{g_2}},g_2}(j')|\sup_{t_1, t_2 \in [0,1]} \left| \gamma_{g_2}(t_1,j')-\gamma_{g_2}(t_2,j') \right|.
\end{split}
\end{align}
Then, there exists some constant $\vartheta = \vartheta(n_1, n_2)>0$, 
\begin{equation*}
\mathbb{P}\Big( |S_{g_1} - S_{g_2}| > \vartheta \Big) = 1-\co(1). 
\end{equation*}  
\end{enumerate}
\end{theorem}

We give the proof in Appendix \ref{append:proofthm2}. We make several remarks. Theorem \ref{thm_2}(a) shows that the maximum deviation feature in (\ref{eq_maxdeviation}) is a consistent estimator of its population counterpart, which encodes the second-order characteristics of the time series. Moreover, the convergence rate can vanish under mild conditions. In particular, for basis functions with $\xi_{c_g} = \cO(1)$ and $\zeta_{c_g} = \cO(\sqrt{c_g})$, when the autocovariance function is smooth enough, i.e., $d_g$ is large, and the time series exhibits short-range dependence, i.e., $\tau_g$ is large, the estimation error is of order $n_g^{-1/2 + \epsilon_g},$ for some small constant $\epsilon_g > 0$.  Theorem \ref{thm_2}(b) shows that, when the two classes of time series differ slightly, in the sense that their second-order characteristics differ in the form of (\ref{eq_assumtiondasdsadasdads}), then the two classes can be effectively differentiated using the aggregated summary  feature in (\ref{maximu_feature_D}).
 
In addition, we make some remarks regarding the technical conditions in this theorem. The condition in (\ref{eq_errorprecisionbound}) ensures the consistency of the estimator $\widehat{\bm{\beta}}$ in \eqref{eq_phihat}. This condition can be easily satisfied for the commonly used basis functions and locally stationary time series models satisfying (\ref{model_1}); see Appendix \ref{append:conditions} for additional discussions. The condition in (\ref{eq_boundoneboundtwo}) ensures that, for relatively large lags, the aggregated summary feature is primarily governed by the underlying second-order characteristics of the time series rather than the estimation error. As a result, when the two classes of time series differ in their second-order characteristics as described in (\ref{eq_assumtiondasdsadasdads}), they can be reliably distinguished. These conditions are satisfied under mild conditions for widely used locally stationary time series models; see Appendix \ref{append:conditions} for additional discussions.

To conclude our theoretical analysis, we next show that our proposed classification procedure is guaranteed to achieve the strong recovery of classification, in that the misclassification rate approaches zero with a high probability. We give the proof in Appendix \ref{append:proofcor1}. 

\begin{corollary}\label{cor_1}
Suppose Assumption \ref{assu_mainmainmain} and (\ref{eq_assumtiondasdsadasdads}) hold for the training time series, and (\ref{eq_boundoneboundtwo}) holds for the training and testing time series. Moreover, suppose all time series are sufficiently long and $M$ in (\ref{equa_tau}) is sufficiently large. Then, 
\begin{equation*} 
\mathbb{P} \left(\widehat{N}_{31} = N_{31}, \widehat{N}_{32} = N_{32} \right) = 1 + \co(1),
\end{equation*} 
where $\widehat{N}_{31}$ and $\widehat{N}_{32}$ denote the number of testing time series in $\{\bm{z}_{k_3}\}_{1 \leq k_3 \leq N_3}$ classified by Algorithm \ref{algo_1} as class 1 and 2, respectively, and $N_{31}$ and $N_{32}$ denote the true number of time series in class 1 and 2, respectively.
\end{corollary}

\section{Simulation Studies} 
\label{sec:simulations}

\subsection{Simulation setup}
\label{subsec:simsetup}

We conduct extensive simulations to investigate the empirical performance of our proposed method. We also numerically compare with the alternative solutions reviewed in Section \ref{subsec:comparison}.

We generate the training time series $\{\bx_{k_1}\}_{1 \leq k_1 \leq N_1}$ and $\{\by_{k_2}\}_{1 \leq k_2 \leq N_2}$ as follows:
\begin{enumerate}[{Model} 1.]
    \item $x_{k_1, i} = 2 \delta \cos(2\pi i / n_{k_1}) x_{k_1, i-1} + \epsilon_{k_1, i}$, \\
    $y_{k_2, i'} = \delta \cos(2\pi i' / m_{k_2}) y_{k_2,i'-1} + \eta_{k_2, i'}$; 
    \item $x_{k_1,i} = 0.4 x_{k_1, i-1} + 0.6 \sin(2\pi i / n_{k_1}) x_{k_1, i-2} + \epsilon_{k_1,i}$, \\ 
    $y_{k_2, i'} = 0.6 y_{k_2, i'-1} + 0.4 \cos(2\pi j / m_{k_2}) y_{k_2, i'-2} + \eta_{k_2, i'}$;
    \item $x_{k_1,i} =  0.4 (\cos(2\pi i / n_{k_1}) + 1) x_{k_1,i-1} + \epsilon_{k_1,i}$, \\ 
    $y_{k_2,i'} = 0.3 \eta_{k_2 ,i'-2} + 0.4 \eta_{k_2 ,i'-1} + \eta_{k_2, i'}$;
    \item $x_{k_1,i} = 1.5 \sin(2\pi i / n_{k_1}) \exp(-i / n_{k_1} x_{k_1 ,i-1}^2) + \epsilon_{k_1,i}$, \\ 
    $y_{k_2 ,i'} = 0.5 \cos(2\pi i' / m_{k_2}) \exp(-i' / m_{k_2} y_{k_2, i'-1}^2) + \eta_{k_2, i'}$;
    \item $x_{k_1,i} = 0.2 \sin(2\pi i / n_{k_1}) x_{k_1, i-1} + 0.2 x_{k_1, i-2} + \epsilon_{k_1,i}$, \\ 
    $y_{k_2, i'} = 0.2 y_{k_2, i'-1} + 0.2 \sin(2\pi i' / m_{k_2}) y_{k_2, i'-2} + \eta_{k_2, i'}$;
    \item $x_{k_1,i} = 0.2 (\sin(2\pi i / n_{k_1}) + 1) / (x_{k_1, i-1} + 1) + 0.2 \exp(-i / n_{k_1} x_{k_1, i-2}^2) + \epsilon_{k_1,i}$, \\ 
    $y_{k_2 ,i'} = 0.2 \exp(-i' / m_{k_2} y_{k_2, i'-1}^2) + 0.3 (\sin(2\pi i' / m_{k_2}) + 1) y_{k_2, i'-2} + \eta_{k_2, i'}$,
\end{enumerate}
where for the white noise terms $\epsilon_{k_1,i}$ and $\eta_{k_2, i'}$, we consider three different forms:
\begin{enumerate}[(i)]
    \item $\epsilon_{k_1,i} = \xi_{k_1,i}, \quad \eta_{k_2, i'} = \xi_{k_2, i'}$; 
    \item $\epsilon_{k_1,i} = \left\{ 1/4 + 1/4 \cos^2(2\pi i/n_{k_1}) \right\} \xi_{k_1,i}, \quad 
    \eta_{k_2, i'} = \left\{ 1/4 + 1/4 \cos^2(2\pi i'/m_{k_2}) \right\} \xi_{k_2, i'}$; 
    \item $\epsilon_{k_1,i} = \left\{ 1/2 + i/(2n_{k_1}) \right\} \xi_{k_1,i}, \quad \eta_{k_2, i'} = \left\{ 1/2 + i'/(2m_{k_2}) \right\} \xi_{k_2, i'}$; 
\end{enumerate}
and $\xi_{k_1,i}, \xi_{k_2, i'}$ are i.i.d.\ standard normal random variables, $i=1, \ldots, n_{k_1}, i'=1, \ldots, m_{k_2}, k_1 = 1, \ldots, N_1, k_2 = 1, \ldots, N_2$. In Model 1, we consider two distinct time-varying AR(1) models, where $\delta$ measures the difference between the two classes of time series, in that the second-order characteristics differ more when $\delta$ increases. We first set $\delta=0.2$. In Models 2 and 5, we consider two time-varying AR(2) models with different AR coefficients. In Model 3, we consider a time-varying AR(1) model and a stationary moving average (MA) model. In Models 4 and 6, we consider two nonlinear time-varying AR(1) and AR(2) models, respectively. We set the length of time series $n = 1000$, and the sample size of the two training groups as $N_1 = N_2 = 100$, and $N_1 = 50, N_2 = 250$.

We also generate the testing time series $\{\bz_{k_3}\}_{1 \leq k_3 \leq N_3}$, with half from class 1 and half from class 2. We set $n =1000$, choose $N_3 = 50$, and apply the classification method to one testing time series in $\{\bz_{k_3}\}_{1 \leq k_3 \leq N_3}$ at a time.

For implementations, our proposed method is available through the $\mathsf{R}$ package $\mathsf{Sie2nts}$, where Legendre polynomials are used as basis functions. The methods, LSW and FLogistic, are implemented using the $\mathsf{R}$ package $\mathsf{wavethresh}$ and $\mathsf{FREG}$, respectively. The methods, MultiRocket, Rocket, Arsenal, STSForest, Shapelet and TimeCNN are implemented using the $\mathsf{Python}$ package $\mathsf{aeon}$, and TimeRNN using $\mathsf{Keras}$. Moreover, we adopt the default architectures, two convolutional layers followed by two pooling layers for TimeCNN, and a three-layer long-short-term-memory for TimeRNN. For the LSW and DWT methods, since the length of the time series must be a power of 2, we set $n = 1024$ for those two methods.

\subsection{Classification accuracy}
\label{subsec:classaccuracy}

\begin{table}[t!]
\centering
\caption{Classification accuracy under the noise distribution (i) for various methods. Reported are the mean and standard deviation (in parentheses) based on 500 data replications.}
\vspace{0.5em}
\label{tab_comparison_1}
\renewcommand{\arraystretch}{1.8}
\setlength{\tabcolsep}{6pt}
\resizebox{\textwidth}{!}{
\begin{tabular}{c|c|c|c|c|c|c|c|c|c|c|c} \hline
\textbf{Method} & Proposed & LSW & DWT & MultiRocket & Rocket & Arsenal & STSForest & Shapelet & FLogistic & TimeCNN & TimeRNN \\
\cline{1-12}
\multicolumn{12}{c}{$N_1 = 100, N_1 = 100$} \\
\cline{1-12}
Model 1 & 0.98 (0.02) & 0.91 (0.04) & 0.75 (0.06) & 0.93 (0.04) & 0.81 (0.06) & 0.81 (0.06) & 0.91 (0.04) & 0.59 (0.07) & 0.50 (0.07) & 0.50 (0.02) & 0.50 (0.07) \\
Model 2 & 0.98 (0.02) & 1.00 (0.00) & 1.00 (0.00) & 1.00 (0.00) & 1.00 (0.01) & 0.99 (0.01) & 0.99 (0.01) & 0.95 (0.04) & 0.50 (0.07) & 0.74 (0.14) & 0.93 (0.05) \\
Model 3 & 1.00 (0.01) & 0.99 (0.01) & 0.86 (0.05) & 1.00 (0.00) & 0.99 (0.01) & 0.99 (0.01) & 0.99 (0.01) & 0.75 (0.08) & 0.51 (0.07) & 0.51 (0.04) & 0.55 (0.07) \\
Model 4 & 0.96 (0.03) & 1.00 (0.00) & 0.99 (0.01) & 1.00 (0.00) & 1.00 (0.00) & 1.00 (0.00) & 1.00 (0.00) & 1.00 (0.00) & 1.00 (0.00) & 1.00 (0.00) & 0.99 (0.04) \\
Model 5 & 0.96 (0.03) & 0.99 (0.00) & 0.99 (0.01) & 1.00 (0.00) & 0.99 (0.01) & 0.99 (0.01) & 0.99 (0.01) & 0.73 (0.08) & 0.50 (0.07) & 0.50 (0.03) & 0.50 (0.07) \\
Model 6 & 0.95 (0.03) & 0.85 (0.07) & 1.00 (0.00) & 0.99 (0.01) & 0.99 (0.01) & 0.99 (0.01) & 0.99 (0.01) & 0.99 (0.01) & 0.47 (0.08) & 0.52 (0.05) & 0.65 (0.07) \\
\cline{1-12}
\multicolumn{12}{c}{$N_1 = 50, N_2 = 250$} \\
\cline{1-12}
Model 1 & 0.95 (0.04) & 0.90 (0.04) & 0.74 (0.06) & 0.84 (0.05) & 0.55 (0.03) & 0.57 (0.04) & 0.81 (0.05) & 0.50 (0.00) & 0.50 (0.07) & 0.50 (0.00) & 0.51 (0.05) \\
Model 2 & 0.96 (0.03) & 0.99 (0.00) & 1.00 (0.00) & 1.00 (0.00) & 0.99 (0.01) & 0.99 (0.01) & 0.99 (0.01) & 0.98 (0.02) & 0.50 (0.07) & 0.61 (0.12) & 0.87 (0.07) \\
Model 3 & 1.00 (0.01) & 0.98 (0.02) & 0.86 (0.05) & 0.99 (0.01) & 0.97 (0.02) & 0.97 (0.02) & 0.99 (0.01) & 0.62 (0.06) & 0.50 (0.07) & 0.50 (0.00) & 0.53 (0.06) \\
Model 4 & 0.96 (0.03) & 1.00 (0.00) & 0.99 (0.01) & 1.00 (0.00) & 1.00 (0.00) & 1.00 (0.00) & 1.00 (0.00) & 1.00 (0.00) & 1.00 (0.00) & 1.00 (0.00) & 0.98 (0.05) \\
Model 5 & 0.96 (0.03) & 0.99 (0.01) & 0.99 (0.00) & 1.00 (0.00) & 0.99 (0.01) & 0.99 (0.01) & 0.99 (0.01) & 0.54 (0.04) & 0.50 (0.07) & 0.50 (0.00) & 0.50 (0.05) \\
Model 6 & 0.95 (0.04) & 0.87 (0.06) & 1.00 (0.00) & 0.99 (0.01) & 0.99 (0.01) & 0.99 (0.01) & 0.99 (0.01) & 0.99 (0.01) & 0.56 (0.03) & 0.50 (0.00) & 0.71 (0.06) \\ \hline
\end{tabular}
}
\end{table}

Table \ref{tab_comparison_1} reports the classification accuracy results under the white noise distribution (i) based on 500 data replications. The results for the white noise distributions (ii) and (iii) are reported in Appendix \ref{append:simulations}. It is seen that our proposed method consistently achieves high accuracy across all settings and outperforms the alternative methods. In particular, for Model 4, all methods perform well. However, when the two classes of time series become more similar, as in Models 1, 3, 5 and 6, FLogistic, TimeRNN, and TimeCNN perform poorly, regardless of whether the data is balanced or not. For Model 1, particularly in the unbalanced case, all methods except for our proposed one and LSW, have the classification accuracy dropping below $85\%$. Finally, LSW performs poorly in the highly nonlinear Model 6.

\begin{figure}[t!]
\centering
\includegraphics[width=1.0\textwidth]{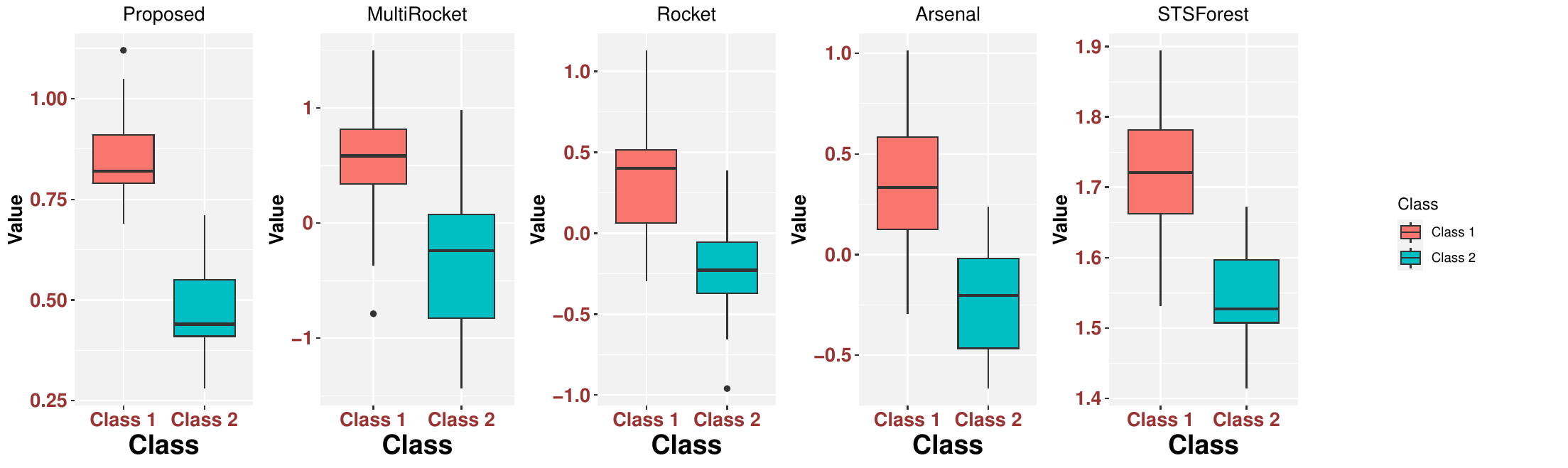}\\
\vspace{1em}
\includegraphics[width=1.0\textwidth]{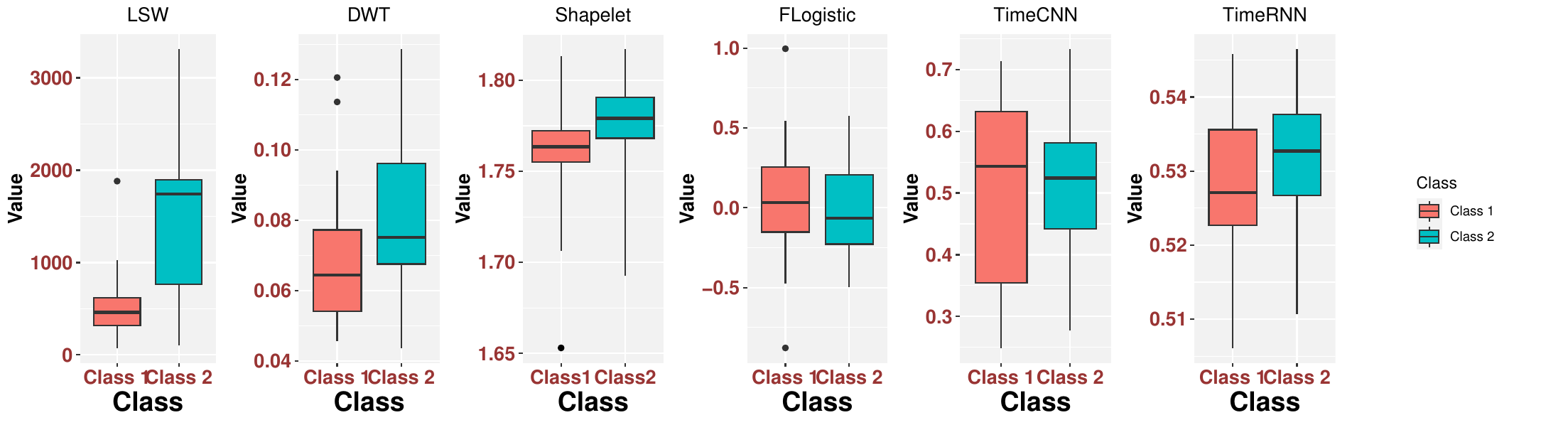}
\caption{Boxplot of extracted features by various methods under Model 1, with $\delta=0.2$, $N_1 = N_2 = 100$, and $n=1000$.}
\label{fig_features}
\end{figure}

Figure \ref{fig_features} reports the extracted features by various methods under Model 1, with $\delta=0.2$, $N_1 = N_2 = 100$, and $n=1000$. It is clearly seen that our extracted feature achieves the most pronounced separation between the two classes among all methods. In contrast, the features derived from the alternative methods show noticeable overlap between the two classes, leading to degraded classification accuracies. 

Intuitively, the reason that our method outperforms the alternative solutions is because our discriminative feature in \eqref{eq_maxdeviation} is designed to capture the difference in the dynamics of locally stationary time series. In contrast, the alternative methods usually rely on simpler features, such as shape-based or frequency-based representations, or basic overall summary statistics derived from the original time series. In conclusion, our method consistently delivers a strong classification performance.

\subsection{Robustness analysis} 
\label{subsec:robust}

We next examine the performance of our proposed method, along with the alternative solutions, when the difference between the two classes becomes more pronounced, when the sample sizes become more imbalanced, and when the length of time series increases. We adopt Model 1 under the white noise distribution (i). 

\begin{figure}[t!]
\centering
\begin{tabular}{ccc}
\includegraphics[width=0.31\textwidth, height=2.25in]{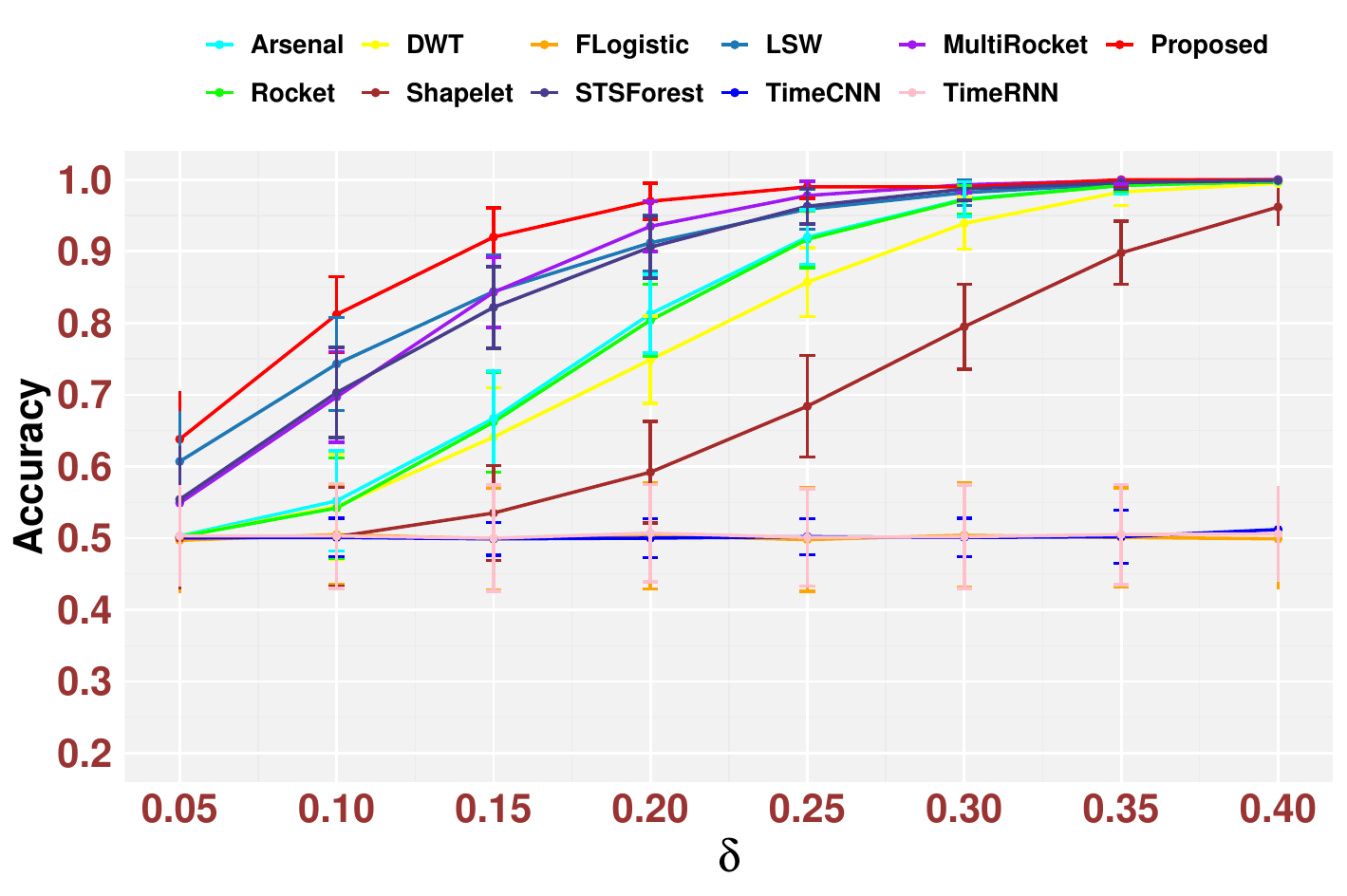} &
\includegraphics[width=0.31\textwidth, height=2.25in]{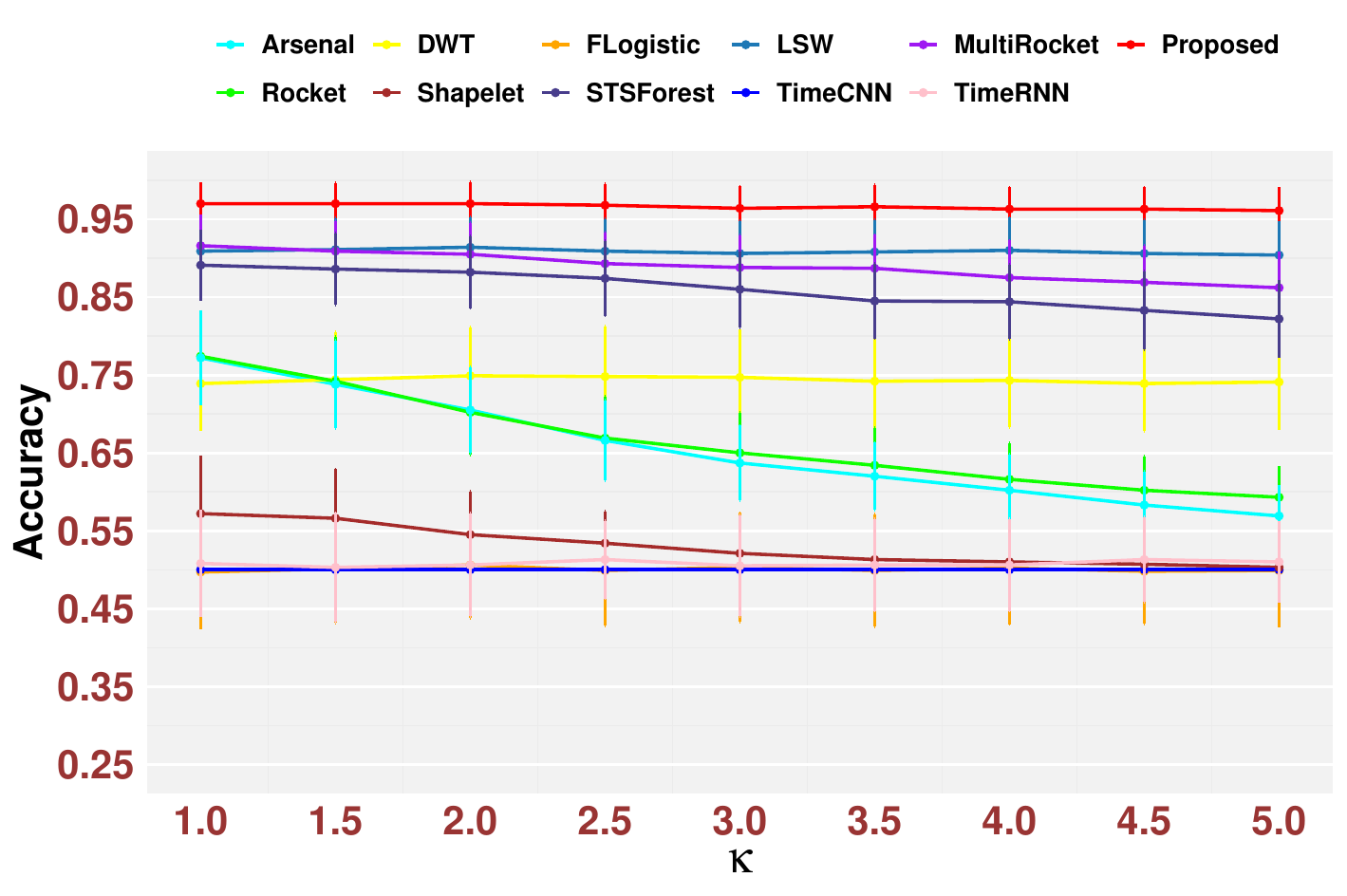} &
\includegraphics[width=0.31\textwidth, height=2.25in]{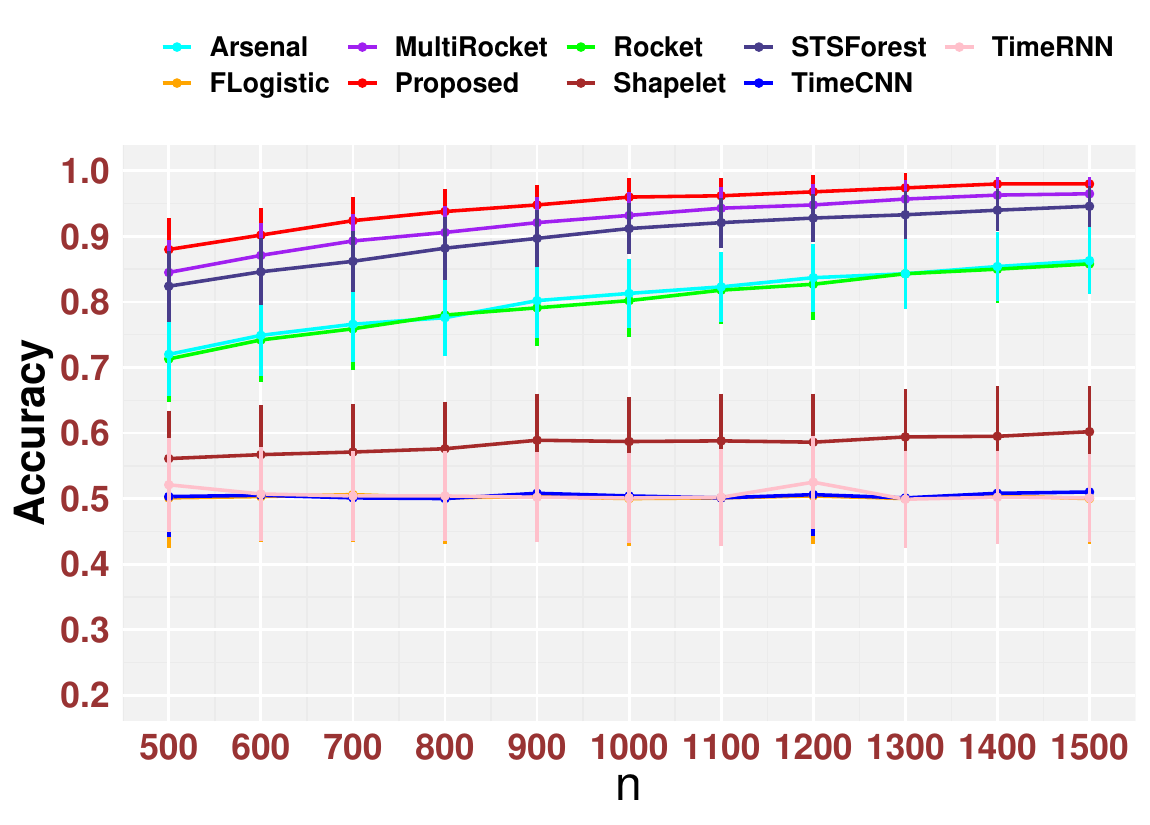} \\
(a) & (b) & (c) 
\end{tabular}
\vspace{-1em}
\caption{Classification accuracy of various methods (a) when the difference between the two classes of time series $\delta$ increases; (b) when the sample size ratio of the two classes $\kappa$ increases; (c) when the length of the time series $n$ increases.}
\label{fig_varying}
\end{figure}

For the difference between the two classes of time series, we vary the value of $\delta$ from 0.05 to 0.4, while we fix $N_1 = N_2 = 100, n=1000$. Figure \ref{fig_varying}(a) reports the results based on 500 data replications. It is seen that our method dominates all other methods, especially when $\delta$ is small, i.e., when the two classes of time series are not clearly separated. Meanwhile, TimeCNN, TimeRNN, and FLogistic consistently show low accuracies, even when $\delta$ increases. 

For the sample size difference between the two classes of time series, we set $N_1 = 50$, $N_2 = \kappa N_1$, and vary the value of $\kappa$ from 1 to 5, while we fix $\delta=0.2, n=1000$. Figure \ref{fig_varying}(b) reports the results based on 500 data replications. It is seen that our proposed method, LSW and DWT maintain high classification accuracy across all values of $\kappa$, demonstrating robustness against imbalanced data, while our method consistently dominates LSW and DWT. Meanwhile, the performances of Shapelet, STSForest, Rocket, MultiRocket and Arsenal decrease as the data becomes more imbalanced, whereas TimeRNN, TimeCNN, and FLogistic consistently perform poorly across all values of $\kappa$. 

Finally, for the length of time series, we vary $n$ from 500 to 1500, while we fix $\delta=0.2, N_1 = N_2 = 100$. For LSW and DWT, we use the closest value to $n$ that is a power of 2. Figure \ref{fig_varying}(c) reports the results based on 500 data replications. It is seen that our method continues to dominate all other methods across all values of $n$. The performances of most methods increase as $n$ increases, while the performances of TimeCNN, TimeRNN, and FLogistic remain poor.

\section{EEG Data Analysis}
\label{sec:realdata}

We revisit the motivating application of EEG-based epilepsy classification. Epilepsy classification using EEG plays a critical role in the clinical diagnosis and management of seizure disorders. EEG provides a non-invasive means to monitor brain activity with millisecond-level precision, enabling detection of abnormal neural dynamics associated with epileptic seizures. The classification task typically involves distinguishing seizure-related EEG patterns from non-seizure activity, a process that has traditionally relied on expert visual inspection. However, manual interpretation is time-consuming, labor-intensive, and prone to subjectivity, prompting growing interest in automated classification techniques. Recent advances in both classical time series analysis and deep learning have shown encouraging results in identifying pathological EEG patterns \citep{Jing2023Development, Kiessner2024Reaching}. Nevertheless, some key obstacles still exist, including the nonstationary nature of EEG signals, significant variability across individuals, low signal-to-noise ratios, and severe class imbalance due to the scarcity of seizure events in many datasets. Developing accurate and robust EEG-based epilepsy classifier for clinical use remains challenging. 

To further evaluate our proposed classification method and compare with the alternative ones reviewed in Section \ref{subsec:comparison}, we employ the Temple University Hospital (TUH) Abnormal EEG Corpus, a widely used benchmark dataset for research in automated EEG analysis. As part of the larger Temple University Hospital EEG database, this corpus provides a collection of clinical EEG recordings labeled as either normal or abnormal based on expert interpretation, and is particularly valuable for developing and evaluating machine learning models aimed at detecting neurological abnormalities such as  epilepsy \citep{obeid2016temple}. EEG signals in the dataset are recorded using multiple scalp electrodes, following standard clinical montages, and capture temporal voltage fluctuations linked to underlying neuronal activity. The abnormal class typically includes a broad spectrum of pathological patterns such as epileptiform discharges, slowing, or other atypical waveforms, while the normal class consists of EEGs without clinically significant findings. The dataset is available at \url{https://isip.piconepress.com/projects/nedc/html/tuh_eeg/}. 

In our data analysis, we focus on the channels, CZ, C3, and C4, that are centrally positioned over the motor and somatosensory cortex. We sample 504 subjects as the training data, and another 150 subjects as the testing data. For both cohorts, half are epileptic, and half are healthy controls. We preprocess the EEG data, by discarding the first minute of recording, and use the remaining up to 20 minutes of recordings. We clip the amplitude values to $\pm 800 \mu V$, and downsample each signal from $50$ Hz to $1$ Hz to speed up the computation. We also adopt the standard practice of normalizing each EEG signal by subtracting its mean and dividing by its standard deviation.

\begin{table}[t!]
\centering
\begin{tabular}{lccccclcccc} \hline
\textbf{Method} & & CZ & C3 & C4 & & & & CZ & C3 & C4 \\ \hline
Proposed     &  & 94.7  & 74.7  & 72.6 & &  \\
MultiRocket & & 78.7  & 74.0  & 76.7  & & FLogistic     & & 58.7  & 52.0  & 54.7  \\
Rocket         & & 79.3  & 70.7  & 68.0  & & DWT           & & 73.3  & 61.3  & 64.7  \\
Arsenal        & & 78.0  & 72.7  & 73.3  & & LSW           & & 64.0  & 63.3  & 58.6  \\
STSForest   &  & 75.3  & 65.3  & 71.3 & & TimeCNN    & & 55.3  & 50.0  & 53.3  \\
Shapelet      &  & 72.7  & 64.0  & 66.7 & & TimeRNN    & & 53.3  & 54.7  & 52.7  \\ \hline
\end{tabular}
\caption{Classification accuracy for TUH EEG data for various methods based on the three channels, CZ, C3, and C4 channels.}
\label{tab_realdata_accuracy}
\end{table}

\begin{figure}[t!]
\centering
\begin{tabular}{ccc}
\includegraphics[width=0.45\textwidth, height=2in]{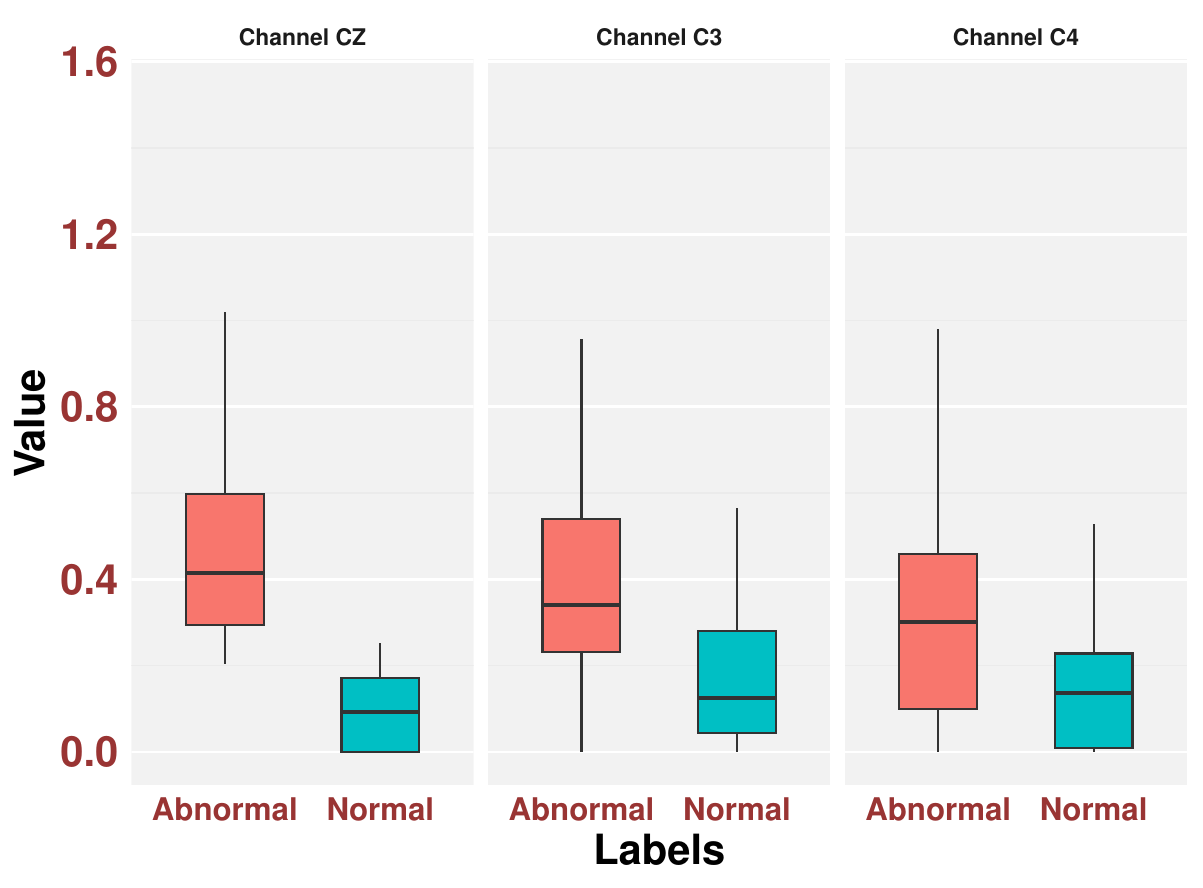} & & 
\includegraphics[width=0.45\textwidth, height=2in]{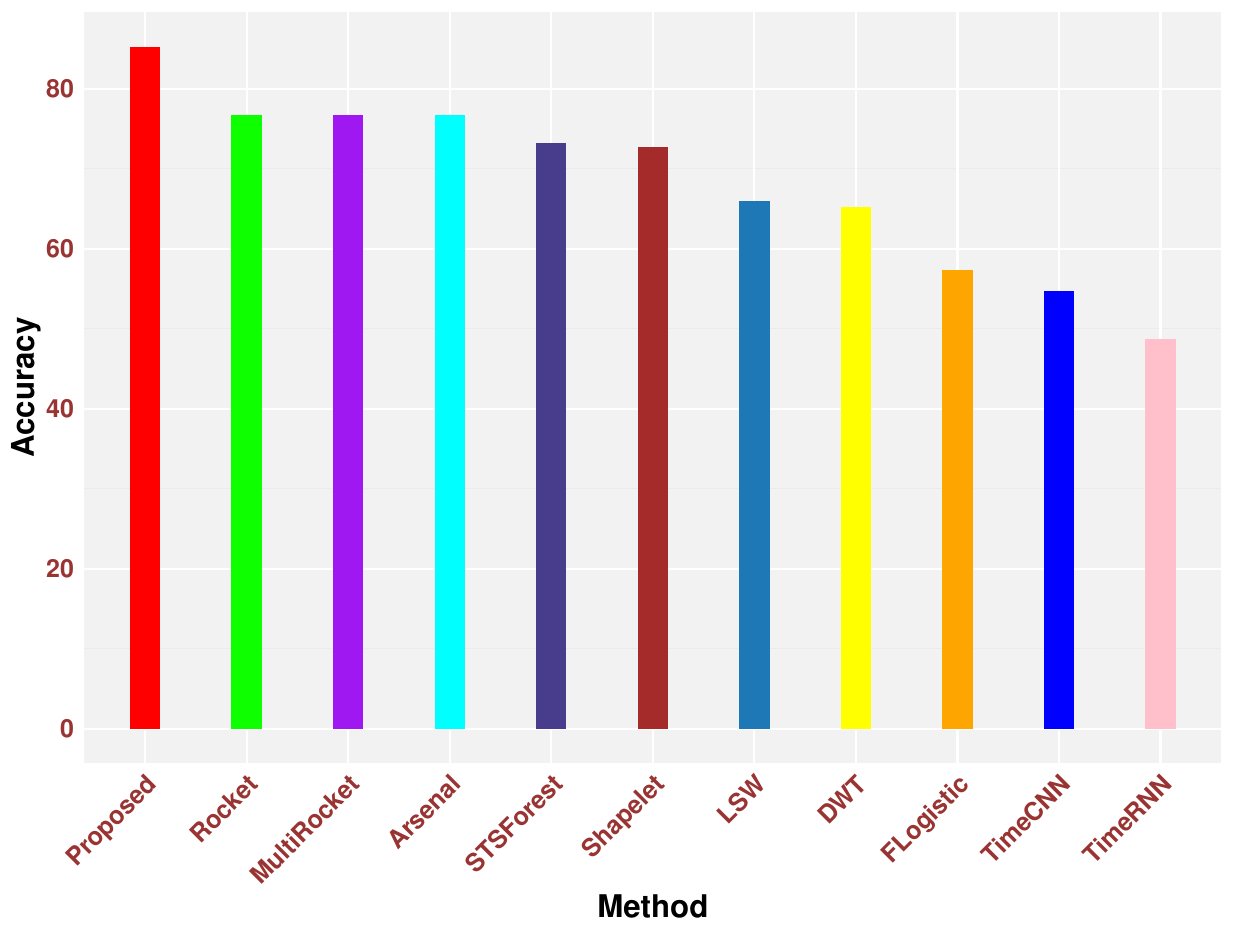} \\
(a) & & (b) 
\end{tabular} 
\caption{Analysis of TUH Abnormal EEG Corpus data. (a) Boxplot of extracted features by various methods for the three EEG channels; (b) Classification accuracy of various methods based on majority voting.}
\label{fig_realdata}
\end{figure}

Table \ref{tab_realdata_accuracy} reports the classification accuracy of various methods based on the three EEG channels independently. It is seen that our proposed method achieves the overall best accuracy, and also the best accuracy for the channels CZ and C3, compared to the alternative solutions. Figure \ref{fig_realdata}(a) reports the extracted features by our proposed method for the two classes for EEG signals by the three channels.  It is seen that the features of the epileptic subjects exhibit higher values than those of health controls across all three channels, reflecting more volatile patterns and greater nonstationarity of the neural dynamics. Figure \ref{fig_realdata}(b) reports classification accuracy of various methods by further adding a majority voting step based on the three channels. It is again seen that our proposed method outperforms all the alternative solutions. 

An interesting direction for future extension of our method is from univariate to multivariate time series. For the EEG classification task, it is of interest to consider multiple channels simultaneously, as abnormal signals may manifest through interactions across different regions of the brain. Incorporating such multivariate information could potentially further improve the classification accuracy and offer deeper physiological insights. However, the multivariate setting introduces additional challenges in terms of structural identification and approximation, such as capturing cross-series dependencies and handling higher-dimensional structures in the time domain. We leave this extension for future investigation.

\renewcommand{\thesubsection}{S\arabic{subsection}}
\renewcommand{\theequation}{S\arabic{equation}}
\renewcommand{\thefigure}{S\arabic{figure}}
\renewcommand{\thetable}{S\arabic{table}}
\renewcommand\thelemma{S\arabic{lemma}}
\renewcommand\theproposition{S\arabic{proposition}}
\renewcommand\thetheorem{S\arabic{theorem}}

\begin{center}
{\Large{\bf Supplementary Appendix for}} \\
\medskip
{\Large{\bf ``Structural Classification of Locally Stationary Time Series Based on Second-order Characteristics''}} \\
\bigskip
\bigskip

{\large{\sc Chen Qian, Xiucai Ding, and Lexin Li}} \\
\bigskip
\end{center}

\baselineskip=19pt


In this supplementary appendix, we first present some additional discussions and simulation results, then present the proofs of all theoretical results in the paper.

\subsection{Additional discussions}

\subsubsection{Classification procedure for correlation stationary time series}
\label{append:stationary}

In this section, we discuss the classification for correlation stationary time series, i.e., when $\phi_j (i/n) \equiv \phi_j$, for $j = 1, \ldots, b$, in \eqref{eq_firstapproximation}. 

We begin with checking the correlation stationarity of an observed time series. We follow the procedure of \cite{ding2023autoregressive}. Recalling (\ref{eq_phihat}), we adopt the following test statistic,
\begin{equation*}
T = \sum_{j=1}^{b} \int_{0}^{1} \left( \widehat \phi_j(t) - \overline{\widehat \phi}_j \right)^2 \mathrm{d}t, \ \overline{\widehat \phi}_j = \int_{0}^{1} \widehat \phi_j(t) \mathrm{d}t. 
\end{equation*}
We compute the $p$-value of this test using the multiplier bootstrap procedure as described in \cite{ding2023autoregressive}. 

Next, we develop a classification procedure when the time series $\bz = (z_i)_{i=1}^{n}$ is correlation stationary. In this case, the AR approximation with order $b$ for $(z_i)_{i=1}^{n}$, $i > b$, is given by 
\begin{equation}\label{D_6}
z_i=\sum_{j=1}^b \phi_j z_{i-j}+\epsilon_i + \co_{L_2}(1).
\end{equation}
We can estimate $\phi_j$'s using OLS. That is, letting $\boldsymbol{\phi} = (\phi_1, \ldots, \phi_b)^\top \in \mathbb{R}^{b}$, we obtain $\boldsymbol{\widehat \phi} = (X^\top X)^{-1}X^\top (z_{b+1}, ..., z_n)^\top$, where $X \in \mathbb{R}^{(n-b) \times b}$ is the ``design matrix" from \eqref{D_6}. When both classes of time series are correlation stationary, we can simply classify the unlabeled time series by examining the constant coefficients at different lags as follows. 

In the first step, we apply the AR approximation \eqref{D_6} to the training data $\{\bx_{k_1}\}_{1 \leq k_1 \leq N_1}$ from class 1, $\{\by_{k_2}\}_{1 \leq k_2 \leq N_2}$ from class 2, and the testing data $\bz$. Denoting $b^x_{k_1}$, $1 \leq k_1 \leq N_1$, $b^y_{k_2}$, $1 \leq k_2 \leq N_2$, and $b^z$ as the selected order, we compute 
\begin{equation}\label{min_b_star}
b_* = \max_{1 \leq k_1 \leq N_1, 1 \leq k_2 \leq N_2} \{b^x_{k_1}, b^y_{k_2}, b^z\}.
\end{equation}
We then employ (\ref{D_6}) with order $b_*$ for all time series. 

In the second step, denoting ${\widehat \phi_{k_1 j}}$, ${\widehat \phi_{k_2 j}}$, ${\widehat \phi_{z j}}$ as the estimated AR approximation coefficients, we compute the corresponding mean estimated AR approximation coefficient as
\begin{equation}\label{D_7}
\bar\phi^x_{j} = \sum_{k_1=1}^{N_1} \frac{1}{N_1} \widehat \phi_{k_1j} , \quad\quad
\bar\phi^y_{j} = \sum_{k_2 = 1}^{N_2} \frac{1}{N_2} \widehat \phi_{k_2 j}, \quad \textrm{ for } 1 \leq j \leq b_*. 
\end{equation}
We then compute the following distance quantifies,
\begin{equation*} \label{D_3}
S_{xz} = \sqrt{\sum_{j=1}^{b_*} (\widehat \phi_{z j} -  \bar \phi^x_{j})^2}, \quad\quad    
S_{yz} = \sqrt{\sum_{j=1}^{b_*} (\widehat \phi_{z j} -  \bar \phi^y_{j})^2}. 
\end{equation*}
We assign $\bz$ to class 1 if $S_{xz} < S_{yz}$, and to class 2 otherwise. 

We remark that using the same $b_*$ for all time series does not affect the effectiveness of our classification procedure, because the quantities $S_{xz}$ and $S_{yz}$ remain close to those obtained using individually selected order for each time series. More specifically, $b_*$ in \eqref{min_b_star} represents the maximum value among all order parameters obtained through the data-driven procedure across different time series. If a time series would have selected a smaller $b < b_*$, approximating it using $b_*$ has negligible impact. This is because the additional AR coefficients at higher lags, i.e., the lags greater than $b$ but smaller than or equal to $b_*$, would be close to zero \citep[][Section 5.4]{montgomery2015introduction}. As a result, $S_{xz}$ and $S_{yz}$ would be dominated by the lower-lag coefficients, those corresponding to the true order $b$, thus preserving the effectiveness of the classification procedure.

\subsubsection{Additional discussions of conditions}
\label{append:conditions}

In this section, we provide additional examples and discussions for the technical conditions in Theorem \ref{thm_2}. More specifically, we first provide a number of common used sieve basis functions and the corresponding bounds for the relevant quantities in \eqref{eq_basicquantity} with these basis functions. Next, we provide a number of locally stationary time series that can be written in the form of \eqref{model_1}, and as such the condition in (\ref{eq_errorprecisionbound}) is satisfied. Finally, we provide a number of examples under which the conditions in (\ref{eq_boundoneboundtwo}) and (\ref{eq_assumtiondasdsadasdads}) are satisfied. In general, these conditions are mild and hold for many commonly encountered examples.

\begin{example}[Examples of commonly used sieve basis functions]
\label{exam_1}
We present several examples of commonly used basis functions $\{\alpha_\ell(t)\}$ used in \eqref{eq_secondapproximation}, along with the corresponding bounds for the relevant quantities in \eqref{eq_basicquantity} for these basis functions. All these examples can be implemented using the $\mathsf{R}$ package $\mathsf{Sie2nts}$.
    
The first example is the set of normalized Legendre polynomials \citep{bell2004special}. The Legendre polynomial of degree $\kappa \in \mathbb{N}$ defined on $[-1,1]$ can be obtained using Rodrigue's formula, 
\begin{equation*}
P_\kappa(t) = \frac{1}{2^\kappa \kappa!} \frac{\mathrm{d}^\kappa}{\mathrm{d}t^\kappa} (t^2 - 1)^n, \quad -1 \leq t \leq 1.
\end{equation*}
In our implementation, we obtain the Legendre polynomials by mapping $[-1,1]$ to $[0,1]$ and then normalizing them. For these polynomials, it is straightforward to verify that for the quantities in \eqref{eq_basicquantity}, we have that $\xi_c = \cO(1)$ and $\zeta_c = \cO(c)$.  

The second example is the set of normalized trigonometric functions, which are given by $\left\{ 1, \sqrt{2} \cos(2 \kappa \pi t), \sqrt{2} \sin(2 \kappa \pi t), \ldots \right\}, \kappa \in \mathbb{N}$. For  (\ref{eq_basicquantity}), we have that $\xi_c = \cO(1)$ and $\zeta_c = \cO(\sqrt{c})$ \citep{chen2007large}.

The third example is the Daubechies wavelet basis \citep{daubechies1992ten}. The Daubechies scaling (father) wavelet function of order $\kappa \in \mathbb{N}$ is a square-integrable function on $\mathbb{R}$, with support on the interval $[0, 2 \kappa - 1]$, and satisfies the recursion equation as,
\begin{equation*}
\varphi(t) = \sqrt{2} \sum_{k=0}^{2\mathsf{\kappa}-1} h_k \varphi(2t - k), \quad \int_{\mathbb{R}} \varphi(t) \, \mathrm{d}t = 1, \quad \int_{\mathbb{R}} \varphi(2t - k)\, \varphi(2t - k') \, \mathrm{d}t = 0,
\end{equation*}
for $k \neq k', k, k' = 0, \ldots, 2\kappa - 1$, where $h_0, h_1, \ldots, h_{2\mathsf{N}-1} \in \mathbb{R}$ are the filter (low-pass) coefficients. For (\ref{eq_basicquantity}), we have that  $\xi_c = \cO(\sqrt{c})$ and $\zeta_c = \cO(\sqrt{c})$. 
\end{example}

\begin{example}[Examples of locally stationary time series satisfying \eqref{model_1}]
\label{exam_2}
We present two examples of locally stationary time series, which encompasses many commonly used models in the literature, that satisfy \eqref{model_1}. For each example, we also illustrate how the physical dependence measure defined in Definition \ref{defn_physcialdependence} can be computed.

The first example is the locally stationary linear process \citep{dahlhaus2019towards,WZ1}. For i.i.d.\ random variables $\{\epsilon_i\}$ and smooth functions $a_j(\cdot)$, $j = 0, 1, \ldots$, defined on $[0, 1]$, consider the linear process,
\begin{equation*}
G(t, \mathcal{F}_i) = \sum_{j = 0}^\infty a_j(t) \epsilon_{i-j}. 
\end{equation*} 
Such a form includes many commonly used models, for instance, time-varying AR, MA, and ARCH models \citep[][Example 2.13]{ding2023autoregressive}. For this process, the physical dependence measure $\delta(j, q)$ in Definition \ref{defn_physcialdependence} satisfies that $\delta(j, q) = \cO\big(\sup_{t \in [0,1]} |a_j(t)|\big)$ \citep{WZ1}.

The second example is a general class of locally stationary nonlinear time series \citep{dahlhaus2019towards,DZ1,WZ1}. For i.i.d.\ random variables ${\epsilon_i}$ and some measurable function $R$, consider the nonlinear process, 
\begin{equation*} 
G\left(t, \mathcal{F}_i\right) = R\big( t, G\left(t, \mathcal{F}_{i-1}\right), \epsilon_i \big).
\end{equation*} 
Such a form includes many commonly used locally stationary nonlinear processes, for instance, the nonlinear time-varying threshold autoregressive model \citep[][Section 2.1]{DZ1}. For this process, the physical dependence measure $\delta(j, q)$ in Definition \ref{defn_physcialdependence} satisfies that $\delta(j, q) = \cO(\chi^j)$ when $\chi < 1$, where 
\begin{equation*} 
\chi = \sup{t \in [0,1]} \sup_{x \neq x'} \frac{\|R(t, x, \epsilon_0) - R(t, x', \epsilon_0)\|_q}{|x - x'|}, 
\end{equation*} 
{for some $x_0$ such that $\sup_{t \in [0,1]} \left| R\left(t, x_0, \epsilon_i \right) \right|_q < \infty$.} 
\end{example}

\begin{example}[Examples of locally stationary time series satisfying (\ref{eq_boundoneboundtwo}) and (\ref{eq_assumtiondasdsadasdads})]
\label{exam_3} 
We present two examples of locally stationary time series that satisfy the conditions in (\ref{eq_boundoneboundtwo}) and (\ref{eq_assumtiondasdsadasdads}). 

The first example includes two classes of time-varying MA(1) models given by 
\begin{eqnarray*}
x_i & = & a_1(i/n_1) \epsilon_{i-1} + \epsilon_i, \;\; \textrm { for } i = 1, \ldots, n_1, \\
y_{i'} & = & a_2(i'/n_2) \eta_{i'-1} + \eta_{i'}, \;\; \textrm { for } i' = 1, \ldots, n_2, 
\end{eqnarray*}
where $a_1(\cdot)$ and $a_2(\cdot)$ are smooth functions, and ${\epsilon_i}$ and $\eta_{i'}$ are i.i.d.\ standard Gaussian random variables. For the two classes of time series, it is straightforward to see that the autocovariance function in (\ref{eqn_covij}) can be written as, 
\begin{eqnarray*}
\gamma_g(t, h) = 
\begin{cases}
a_g^2(t) + 1, & h = 0, \\
a_g(t), & h = 1, \\
0, & h \ge 2,
\end{cases}
\ \ g=1,2, 
\end{eqnarray*}
when $h \geq 0$. Consequently, we see that, for $g=1,2$, when the lag $j \asymp b_g^{a_g}$ with $0<a_g \leq 1$, Theorem \ref{thm_1} implies that, when $\sup_{t_1, t_2}|a_g(t_1)-a_g(t_2)| \gg n_g^{-C_g}$ for some large positive constants $C_1$ and $C_2$, we have $D_g^*(j) \asymp \sup_{t_1, t_2}|a_g(t_1)-a_g(t_2)|$. Together with Theorem \ref{thm_2}, we see that our proposed algorithm is to achieve a consistent classification performance if, for some large constant $C$, $\sup_{t_1, t_2}|a_1(t_1)-a_1(t_2)| > C \sup_{t_1, t_2}|a_2(t_1)-a_2(t_2)|$, or vice versa.

The second example includes two classes of time-varying AR(1) models given by 
\begin{eqnarray*}
x_i & = & a_1(i/n_1)x_{i-1} + \epsilon_i, \;\; \textrm { for } i = 1, \ldots, n_1, \\
y_{i'} & = & a_2(i'/n_2) y_{i'-1} + \eta_{i'}, \;\; \textrm { for } i' = 1, \ldots, n_2,
\end{eqnarray*}
where $a_1(\cdot)$ and $a_2(\cdot)$ are smooth functions satisfying that $\sup_{t \in [0,1]} |a_1(t)| < 1$ and $\sup_{t \in [0,1]} |a_2(t)| < 1$, and ${\epsilon_i}$ and ${\eta_{i'}}$ are i.i.d.\ standard Gaussian random variables. Following \cite{zhou2013inference}, the autocovariance functions for $h \geq 0$ satisfy that,
\begin{align*}
\gamma_g(t, h) = \frac{a_g^h(t)}{1 - a_g^2(t)} + \mathrm{O}(\log^2(n_g)/n_g), \ g=1,2. 
\end{align*}
Denote $\psi_g(t) = a_g(t)/[\{1-a_g^2(t)\}\{1-a_g(t)\}]$. Then, for $g=1,2$, when the lag $j \asymp b_g^{a_g}$ with $0<a_g \leq 1$, Theorem \ref{thm_1} implies that, when $\sup_{t_1, t_2}|\psi_g(t_1)-\psi_g(t_2)| \gg n_g^{-C_g}$ for some large positive constants $C_1$ and $C_2$, we have $D_g^*(j) \asymp \sup_{t_1, t_2}|\psi_g(t_1)-\psi_g(t_2)| \asymp \sup_{t_1, t_2}|a_g(t_1)-a_g(t_2)|$. Together with Theorem \ref{thm_2}, we see that our proposed algorithm is to achieve a consistent classification performance if, for some large constant $C$, $\sup_{t_1, t_2}|a_1(t_1)-a_1(t_2)| > C \sup_{t_1, t_2}|a_2(t_1)-a_2(t_2)|$, or vice versa.
\end{example}

\subsection{Additional simulation results}
\label{append:simulations}

\begin{table}[t!]
\centering
\caption{Classification accuracy under the noise distribution (ii) for various methods. Reported are the mean and standard deviation (in parentheses) based on 500 data replications.}
\vspace{0.5em}
\label{tab:comparison_2}
\renewcommand{\arraystretch}{1.8}
\setlength{\tabcolsep}{6pt}
\resizebox{\textwidth}{!}{
\begin{tabular}{c|c|c|c|c|c|c|c|c|c|c|c} \hline
\textbf{Method} & Proposed & LSW & DWT & MultiRocket & Rocket & Arsenal & STSForest & Shapelet & FLogistic & TimeCNN & TimeRNN \\ 
\cline{1-12}
\multicolumn{12}{c}{$N_1 = 100, N_1 = 100$} \\
\cline{1-12}
        Model 1 & 0.95 (0.03) & 0.90 (0.04) & 0.78 (0.06) & 0.93 (0.03) & 0.81 (0.05) & 0.81 (0.06) & 0.90 (0.04) & 0.56 (0.07) & 0.50 (0.07) & 0.55 (0.07) & 0.50 (0.07) \\
        Model 2 & 0.96 (0.03) & 1.00 (0.00) & 1.00 (0.00) & 1.00 (0.00) & 0.99 (0.01) & 0.99 (0.00) & 0.99 (0.00) & 0.93 (0.04) & 0.50 (0.07) & 0.59 (0.12) & 0.95 (0.03) \\
        Model 3 & 0.99 (0.01) & 0.99 (0.01) & 0.89 (0.04) & 0.99 (0.00) & 0.99 (0.01) & 0.99 (0.01) & 0.99 (0.01) & 0.83 (0.07) & 0.52 (0.07) & 0.50 (0.02) & 0.52 (0.07) \\
        Model 4 & 0.99 (0.01) & 1.00 (0.00) & 1.00 (0.00) & 1.00 (0.00) & 1.00 (0.00) & 1.00 (0.00) & 1.00 (0.00) & 1.00 (0.00) & 1.00 (0.00) & 1.00 (0.00) & 0.99 (0.03) \\
        Model 5 & 0.97 (0.03) & 0.99 (0.00) & 0.99 (0.00) & 1.00 (0.00) & 0.99 (0.00) & 0.99 (0.00) & 0.99 (0.01) & 0.70 (0.08) & 0.50 (0.08) & 0.50 (0.07) & 0.50 (0.07) \\
        Model 6 & 0.95 (0.03) & 0.97 (0.02) & 0.78 (0.07) & 0.99 (0.00) & 0.99 (0.01) & 0.99 (0.00) & 0.99 (0.01) & 0.78 (0.07) & 0.79 (0.07) & 0.54 (0.10) & 0.76 (0.06) \\
\cline{1-12}
\multicolumn{12}{c}{$N_1 = 50, N_2 = 250$} \\
\cline{1-12}
        Model 1 & 0.95 (0.03) & 0.88 (0.04) & 0.77 (0.06) & 0.86 (0.05) & 0.59 (0.04) & 0.56 (0.04) & 0.82 (0.05) & 0.50 (0.01) & 0.50 (0.07) & 0.50 (0.00) & 0.51 (0.06) \\
        Model 2 & 0.95 (0.03) & 0.99 (0.01) & 1.00 (0.00) & 1.00 (0.00) & 0.99 (0.00) & 0.99 (0.00) & 0.99 (0.00) & 0.89 (0.07) & 0.50 (0.07) & 0.51 (0.05) & 0.90 (0.07) \\
        Model 3 & 1.00 (0.01) & 0.98 (0.01) & 0.89 (0.04) & 0.99 (0.00) & 0.98 (0.02) & 0.97 (0.02) & 0.99 (0.01) & 0.72 (0.08) & 0.51 (0.07) & 0.50 (0.00) & 0.55 (0.06) \\
        Model 4 & 1.00 (0.01) & 1.00 (0.00) & 1.00 (0.00) & 1.00 (0.00) & 1.00 (0.00) & 1.00 (0.00) & 1.00 (0.00) & 1.00 (0.00) & 1.00 (0.00) & 1.00 (0.00) & 0.99 (0.06) \\
        Model 5 & 0.95 (0.03) & 0.99 (0.01) & 0.99 (0.00) & 0.99 (0.01) & 0.99 (0.01) & 0.99 (0.01) & 0.99 (0.01) & 0.54 (0.05) & 0.50 (0.07) & 0.50 (0.00) & 0.50 (0.06) \\
        Model 6 & 0.95 (0.03) & 0.98 (0.02) & 0.83 (0.09) & 0.99 (0.01) & 0.99 (0.01) & 0.99 (0.01) & 0.99 (0.01) & 0.69 (0.08) & 0.68 (0.08) & 0.50 (0.00) & 0.78 (0.07) \\ \hline
\end{tabular}
}
\end{table}

In this section, we present additional results on classification accuracy for the simulation models under the white noise distribution (ii) and (iii) in Section \ref{subsec:simsetup}. The rest of the setup is the same as before. Specifically, Table \ref{tab:comparison_2} reports the results for the noise distribution (ii), and Table \ref{tab:comparison_3} for (iii). We observe the same qualitative patterns as before, and our proposed classification method continues to performs the best among all solutions. 
 
\begin{table}[t!]
\centering
\caption{Classification accuracy under the noise distribution (iii) for various methods. Reported are the mean and standard deviation (in parentheses) based on 500 data replications.}
\vspace{0.5em}
\label{tab:comparison_3}
\renewcommand{\arraystretch}{1.8}
\setlength{\tabcolsep}{6pt}
\resizebox{\textwidth}{!}{
\begin{tabular}{c|c|c|c|c|c|c|c|c|c|c|c} \hline
\textbf{Method} & Proposed & LSW & DWT & MultiRocket & Rocket & Arsenal & STSForest & Shapelet & FLogistic & TimeCNN & TimeRNN \\
\cline{1-12}
\multicolumn{12}{c}{$N_1 = 100, N_1 = 100$} \\
\cline{1-12}
            Model 1 & 0.96 (0.03) & 0.91 (0.04) & 0.76 (0.06) & 0.93 (0.03) & 0.77 (0.06) & 0.78 (0.05) & 0.90 (0.04) & 0.52 (0.07) & 0.49 (0.07) & 0.55 (0.08) & 0.53 (0.07) \\
            Model 2 & 0.97 (0.02) & 0.99 (0.00) & 1.00 (0.00) & 1.00 (0.00) & 0.99 (0.00) & 0.99 (0.00) & 0.99 (0.00) & 0.96 (0.04) & 0.50 (0.07) & 0.70 (0.15) & 0.97 (0.04) \\
            Model 3 & 1.00 (0.01) & 0.99 (0.01) & 0.89 (0.05) & 1.00 (0.00) & 0.98 (0.02) & 0.99 (0.01) & 0.99 (0.01) & 0.75 (0.08) & 0.51 (0.07) & 0.50 (0.03) & 0.58 (0.07) \\
            Model 4 & 0.97 (0.02) & 1.00 (0.00) & 0.99 (0.00) & 1.00 (0.00) & 1.00 (0.00) & 1.00 (0.00) & 1.00 (0.00) & 1.00 (0.00) & 1.00 (0.00) & 1.00 (0.00) & 0.99 (0.03) \\
            Model 5 & 0.97 (0.02) & 0.99 (0.00) & 0.99 (0.00) & 1.00 (0.00) & 0.99 (0.00) & 1.00 (0.01) & 1.00 (0.00) & 0.70 (0.07) & 0.50 (0.07) & 0.50 (0.07) & 0.50 (0.07) \\
            Model 6 & 0.95 (0.03) & 0.85 (0.07) & 0.99 (0.01) & 0.99 (0.01) & 0.99 (0.01) & 0.99 (0.01) & 0.99 (0.01) & 0.98 (0.02) & 0.49 (0.06) & 0.57 (0.11) & 0.62 (0.07) \\
\cline{1-12}
\multicolumn{12}{c}{$N_1 = 50, N_2 = 250$} \\
\cline{1-12}
            Model 1 & 0.94 (0.04) & 0.90 (0.04) & 0.74 (0.06) & 0.86 (0.04) & 0.53 (0.03) & 0.52 (0.02) & 0.80 (0.05) & 0.50 (0.01) & 0.50 (0.08) & 0.52 (0.05) & 0.48 (0.05) \\
            Model 2 & 0.96 (0.03) & 0.99 (0.00) & 1.00 (0.00) & 1.00 (0.00) & 1.00 (0.00) & 0.99 (0.01) & 0.99 (0.01) & 0.93 (0.06) & 0.50 (0.07) & 0.57 (0.10) & 0.90 (0.07) \\
            Model 3 & 1.00 (0.01) & 0.98 (0.02) & 0.88 (0.05) & 0.99 (0.00) & 0.97 (0.03) & 0.96 (0.03) & 0.99 (0.01) & 0.63 (0.08) & 0.51 (0.07) & 0.50 (0.00) & 0.56 (0.06) \\
            Model 4 & 0.96 (0.03) & 1.00 (0.00) & 0.99 (0.01) & 1.00 (0.00) & 1.00 (0.00) & 1.00 (0.00) & 1.00 (0.00) & 1.00 (0.00) & 1.00 (0.00) & 1.00 (0.00) & 0.99 (0.06) \\
            Model 5 & 0.96 (0.03) & 0.99 (0.01) & 1.00 (0.00) & 0.99 (0.00) & 0.99 (0.00) & 0.99 (0.01) & 0.99 (0.00) & 0.53 (0.03) & 0.50 (0.07) & 0.50 (0.05) & 0.50 (0.05) \\
            Model 6 & 0.94 (0.04) & 0.87 (0.07) & 0.99 (0.01) & 0.99 (0.01) & 0.98 (0.02) & 0.98 (0.02) & 0.99 (0.01) & 0.98 (0.02) & 0.54 (0.03) & 0.50 (0.00) & 0.64 (0.06) \\ \hline
\end{tabular}
}
\end{table}

\subsection{Proofs}
\label{append:proofs}

In this section, we provide the technical proofs of the main results presented in Section \ref{sec:theory}. 

We first summarize some important results from the literature. According to Lemma 2.6 of \cite{DZ1}, under Assumption \ref{assu_basicassumption}, for the autocovariance function, we have,
\begin{equation}\label{eq_autocovcontrol}
\sup_t|\gamma(t,j)|=\cO(j^{-\tau}), \ j \geq 1. 
\end{equation} 
Moreover, according to the proof of Theorem 2.4 of \cite{ding2023autoregressive}, for $b>0$ and the $b \times b$ matrix $\Gamma^b(t)=(\Gamma^b_{\ell_1 \ell_2}(t)) \in \mathbb{R}^{b \times b}$ whose entries are defined as in (\ref{eq_gammaentries}) with $j=b,$ we have that, for its inverse matrix $\Omega^b(t)=(\Gamma^b(t))^{-1}$ in (\ref{eq_precision})
\begin{equation}\label{eq_inversebound}
\sup_t|\Omega^b_{\ell_1 \ell_2}(t)|=\cO\left( (b/\log b)^{-\tau+1}+ \mathbb{I}(|\ell_1-\ell_2|<b/\log b) r^{|\ell_1-\ell_2| \log b/b} \right),
\end{equation}
where $0<r<1$ is some universal constant.

\subsubsection{Proof of Theorem \ref{thm_1}}
\label{append:proofthm1}

For part (a), given $b>0$, for (\ref{eq_firstapproximation}), write $\boldsymbol{\phi}(t) = (\phi_{1}(t), \ldots, \phi_{b}(t))^\top \in \mathbb{R}^b$. Recall that $\Omega^b(t) = (\Gamma^b(t))^{-1}$ in (\ref{eq_precision}). By Theorem 2.11 of \cite{ding2023autoregressive}, we have that  
\begin{equation}\label{yule}
\boldsymbol{\phi}(t) = \Omega^b(t) \bm{\nu}^b(t),
\end{equation} 
where $\boldsymbol{\nu}^b(t)=(\nu^b_1(t), \ldots, \nu^b_b(t))^\top \in \mathbb{R}^b$, with $\nu^b_\ell(t)=\gamma(t,\ell), 1 \leq \ell \leq b.$ By Assumption \ref{assu_basicassumption}, (\ref{yule}) is well defined for all $b$. Therefore, part (a) of the theorem comes directly from (\ref{yule}). 

For part (b), the first result, using (\ref{yule}), for $1 \leq j \leq b$, we have
\begin{align}\label{eq_basicdecomposition}
\phi_j(t_1)-\phi_j(t_2)& =\left[\Omega^b(t_1) \bm{\nu}^b(t_1)-\Omega^b(t_1) \bm{\nu}^b(t_2) \right]^\top \bm{e}_j+\left[\Omega^b(t_1) \bm{\nu}^b(t_2)-\Omega^b(t_2) \bm{\nu}^b(t_2)\right]^\top \bm{e}_j \nonumber \\
&:=P_1 + P_2.
\end{align} 
From (\ref{eq_basicdecomposition}), we have that
\begin{equation*}
|\phi_j(t_1)-\phi_j(t_2)| \leq |P_1| + |P_2|.
\end{equation*} 

For $P_1,$ by definition, we have that 
\begin{equation*}
P_1=\sum_{j'=1}^b (\gamma(t_1,j')-\gamma(t_2,j')) \Omega^b_{j'j}(t_1).
\end{equation*}
By the definition (\ref{eq_wjkdefinition}) and the bound in (\ref{eq_inversebound}), we see from triangle inequality that 
\begin{equation*}
\sup_{t_1, t_2 \in [0,1]}|P_1|=\cO\left( \sum_{j'=1}^b w_j(j') \sup_{t_1, t_2 \in [0,1]} |\gamma(t_1, j')-\gamma(t_2,j')| \right).
\end{equation*}

For $P_2,$ we have that
\begin{equation*}
P_2=\sum_{j'=1}^b  \gamma(t_2,j')(\Omega^b_{j'j}(t_1)-\Omega^b_{j'j}(t_2)). 
\end{equation*}
Together with (\ref{eq_autocovcontrol}), we have that 
\begin{equation}\label{eq_P2RESULS}
\sup_{t_1, t_2 \in [0,1]}|P_2|=\cO\left( \sum_{j'=1}^b (j')^{-\tau} \sup_{t_1, t_2 \in [0,1]} |\Omega^b_{j'j}(t_1)-\Omega^b_{j'j}(t_2)| \right).
\end{equation}
This leads to the first result of part (b). 

For part (b), the second result, when $b$ is large, we have that
\begin{align*}
\sum_{j'=1}^b (j')^{-\tau} r^{|j'-j| \log b/b} & =\sum_{j'=1}^{j-1} (j')^{-\tau} r^{(j-j') \log b/b}+\sum_{j'=j}^b (j')^{-\tau} r^{(j'-j) \log b/b} \\
& =\sum_{j'=1}^{j-1} (j')^{-\tau} r^{(j-j') \log b/b}+\cO\left(j^{-\tau+1} \right). 
\end{align*}
Since $j \asymp b^{a}, 0<a\leq 1,$ we have that 
\begin{equation*}
\sum_{j'=1}^{j-1} (j')^{-\tau} r^{(j-j') \log b/b}=\cO \left(\sum_{j'=1}^{j-1} r^{(j-j') \log j/j} \right).
\end{equation*}
Following a similar discussion of (D.8) in \cite{ding2023autoregressive}, we have that 
\begin{equation*}
\sum_{j'=1}^{j-1} r^{(j-j') \log j/j}=\cO\left( j^{-\tau+1}\right). 
\end{equation*}
Combining the above results with (\ref{eq_inversebound}) and (\ref{eq_P2RESULS}) leads to the second result of part (b). 

For part (c), using (\ref{eq_basicdecomposition}), we have that 
\begin{equation*}
D^*(j) \geq  | P_1 |-|P_2|.
\end{equation*} 
Then the rest of the proof comes from the second result of part (b). 

Together, this completes the proof of Theorem \ref{thm_1}. 
\eop

\subsubsection{Proof of Theorem \ref{thm_2}}
\label{append:proofthm2}

For notional simplicity, denote $\phi_{j,1}(t), 1 \leq j \leq b_1$, and $\phi_{j,2}(t), 1 \leq j \leq b_2$, as the smooth AR approximation coefficient functions for the two classes of time series,  respectively. By Theorem 2.11 of \cite{ding2023autoregressive}, under Assumption \ref{assu_mainmainmain}, we have that $\phi_{j,g}(t)\in C^{d_g}([0, 1])$, $1 \le j \le b_g$, $g=1,2$. 

For part (a), without loss of generality, we focus our proof for $D_1$. Denote
\begin{equation*}
\phi_{j,1,c_1}=\sum_{\ell=1}^{c_1} a_{j\ell,1} \alpha_\ell(t).  
\end{equation*}
Let $\widehat{\phi}_{j,1}(t)$ be the sample estimator as in (\ref{eq_phihat}). By triangle inequality, we have 
\begin{equation*}
\sup_j\|\phi_{j,1}(t)-\widehat{\phi}_{j,1}\| \leq \sup_j |\phi_{j,1}-\phi_{j,1,c_1}(t)| + \sup_j \|\phi_{j,1,c_1}(t)-\widehat{\phi}_{j,1}(t)\|.
\end{equation*} 
Note that $\sup_j |\phi_{j,1}-\phi_{j,1,c_1}(t)|$ can be bounded using (\ref{eq_secondapproximation}), in that 
\begin{equation*}
\sup_j |\phi_{j,1}-\phi_{j,1,c_1}(t)|=\cO\left(c_1^{-d_1} \right).
\end{equation*}
Meanwhile, for $\sup_j \|\phi_{j,1,c_1}(t)-\widehat{\phi}_{j,1}(t)\|$, by (\ref{eq_third_ols}) and OLS, using (\ref{eq_phihat}) and $\bm{\beta} \in \mathbb{R}^{b_1c_1}$ collecting all $\{a_{j\ell,1}\}$, we have that 
\begin{equation*}
\phi_{j,1,c_1}(t)-\widehat{\phi}_{j,1}(t)=\left(\bm{\beta}-\widehat{\bm{\beta}} \right)^\top \mathbf{B}_j(t). 
\end{equation*} 
Consequently, by the definition (\ref{eq_basicquantity}), we have that 
\begin{equation}\label{eq_deodeodeodoeodeoeo}
\sup_j \|\phi_{j,1,c_1}(t)-\widehat{\phi}_{j,1}(t) \| \leq \| \bm{\beta}-\widehat{\bm{\beta}} \| \zeta_{c_1}. 
 \end{equation}
Thus it suffices to bound $\bm{\beta}-\widehat{\bm{\beta}}$. 

Toward that end, we have that  
 \begin{equation*}
 \bm{\beta}-\widehat{\bm{\beta}}=(n_1^{-1}Y^\top Y)^{-1} \frac{Y^\top \bm{\epsilon}}{n_1},
 \end{equation*}
where $\bm{\epsilon}=(\epsilon_{b_1+1} + \tilde{e}, \ldots, \epsilon_{n_1} + \tilde{e})$ with $\tilde{e}$ representing the error term in (\ref{eq_third_ols}). Following a similar idea of \cite{ding2024partial}, we have that 
\begin{equation*}
\left\| \frac{1}{n_1} Y^\top Y- \int_0^{1} \Gamma^{b_1}(t) \otimes \left( \bm{\alpha}(t) \bm{\alpha}(t)^\top \right)  \right\|=\cO\left(b_1c_1 \left(\frac{\xi_{c_1}^2}{\sqrt{n_1}}+\frac{\xi_{c_1}^2 n_1^{\frac{2}{\tau_1+1}}}{n_1} \right) \right),
\end{equation*} 
where $\Gamma^{b_1}(t) \in \mathbb{R}^{b_1 \times b_1}$ is defined as in (\ref{eq_gammaentries}), $\bm{\alpha}(t) \in \mathbb{R}^{c_1}$ collects the basis functions, and $\otimes$ is the Kronecker product. By assumption (\ref{eq_errorprecisionbound}) and Lemma A.1 of \cite{ding2024partial}, we obtain that
\begin{equation*}
\left\| (n_1^{-1}Y^\top Y)^{-1} \right\|=\cO_{\mathbb{P}}(1).
\end{equation*} 
Finally, following a similar discussion in \cite{ding2024partial}, we have that 
\begin{equation*}
\left\| \frac{Y^\top \bm{\epsilon}}{n_1} \right\|=\cO_{\mathbb{P}}\left( \xi_{c_1} \left(1+\frac{b_1^2}{n_1}+b_1 c_1^{-d_1} \right) \sqrt{\frac{b_1c_1}{n_1}} \right). 
\end{equation*} 
Combining the above bounds, we obtain that
\begin{equation*}
\|\bm{\beta}-\widehat{\bm{\beta}} \|=\cO_{\mathbb{P}} \left(\xi_{c_1} \left(1+\frac{b_1^2}{n_1}+b_1 c_1^{-d_1} \right) \sqrt{\frac{b_1c_1}{n_1}}\right). 
\end{equation*}
Combining this with (\ref{eq_deodeodeodoeodeoeo}) leads to part (a). 

For part (b), without loss of generality, suppose that $g_1=1$ and $g_2=2$. By part (a) of this theorem, part (c) of Theorem \ref{thm_1}, and by assumption (\ref{eq_boundoneboundtwo}), we have that, for some constant $c>0$,
\begin{equation*}
S_1>c \sup_{\max\{b_1-b_1^*+1, b_1^*\} \leq j \leq b_1} \sup_{t_1, t_2 \in [0,1]}\left|\sum_{j'=1}^b w_j(j')(\gamma_1(t_1,j')-\gamma_1(t_2,j')) \right| \left(1+\co_{\mathbb{P}}(1) \right). 
\end{equation*}
On the other hand, by part (a) of this theorem, part (b) of Theorem \ref{thm_1}, and assumptions (\ref{eq_boundoneboundtwo}) and (\ref{eq_assumtiondasdsadasdads}), we have that 
\begin{equation*}
S_2 \leq C^{-1}  \sup_{\max\{b_1-b_1^*+1, b_1^*\} \leq j \leq b_1} \sup_{t_1, t_2 \in [0,1]}\left|\sum_{j'=1}^b w_j(j')(\gamma_1(t_1,j')-\gamma_1(t_2,j')) \right| \left(1+\co_{\mathbb{P}}(1) \right).
\end{equation*}
Since the constant $C$ in (\ref{eq_assumtiondasdsadasdads}) is sufficiently large, we therefore have,  
\begin{equation*}
\mathbb{P}(S_1>S_2)=1+\co(1). 
\end{equation*}

This completes the proof of Theorem \ref{thm_2}. 
\eop

\subsubsection{Proof of Corollary \ref{cor_1}}
\label{append:proofcor1}

Since our algorithm does not rely on the number of time series $N_1, N_2$ and $N_3$, in what follows, without loss of generality, we suppose $N_1=N_2=N_3=1$. Denote $S_g^*$ as in (\ref{eq_featurefinaldefinition}) by replacing $D_g(j)$ with $D_g^*(j)$ in (\ref{eq_populationfeatures}). Following the proof of Theorem \ref{thm_2}, under assumption (\ref{eq_assumtiondasdsadasdads}), and without loss of generality, suppose $g_1=1$ and $g_2=2$, we have that 
\begin{equation}\label{eq_consitency}
S_g=S^*_g(1+\co(1)), \ g=1,2, \ S_1^*-S_2^*=(S_1-S_2)(1+\co(1)), \ \text{and} \ S_1^*>S_2^*. 
\end{equation} 
Let $S_3$ and $S_3^*$ denote the associated quantities with the testing data $\bm{z}$. Theoretically, for any threshold $\vartheta \in [S_2^*, S_1^*]$, $\mathbf{z}$ is from class 1 if $S^*_3 \geq \vartheta$, and from class 2 otherwise. Moreover, by part (a) of Theorem \ref{thm_2} and (\ref{eq_consitency}), with a high probability, Algorithm \ref{algo_1} is to assign $\bm{z}$ correctly to class 1 if $\vartheta \in [S_2, S_1]$ and $S_3 \geq \vartheta$.

Finally, based on our choice in (\ref{com_C}), we have that, with a high probability and when the time series are sufficiently long,
\begin{equation*} 
{C}_1 < S_2 < S_1 < {C}_2. 
\end{equation*}
Consequently, when $M$ is sufficiently large, some of the grid points in $\vartheta_i$ would, with a high probability, fall within the interval $[S_2, S_1]$. Moreover, the procedure described in Section \ref{subsec:classify} would select one of these thresholds that yields the best practical performance and again, with a high probability, assigns $\bm{z}$ to the correct class. 
 
This completes the proof of Corollary \ref{cor_1}.
\eop

\bibliographystyle{apalike}
\bibliography{ref-ts}

\end{document}